\documentclass[12pt]{article}
\usepackage{amsmath}
\usepackage{graphicx}

\usepackage{enumerate}
\usepackage{natbib}
\usepackage{url} 

\newcommand{\blind}{1}

\usepackage{verbatim,color,amssymb}
\usepackage{amsmath}					
\usepackage{amsthm}					
\usepackage{natbib}
\usepackage{multirow}
\usepackage{setspace}
\usepackage[mathscr]{euscript}
\usepackage{fancyhdr}
\usepackage{enumitem}
\usepackage{graphicx}
\usepackage{lineno}
\usepackage{array,booktabs}
\usepackage{comment}

\usepackage{multibib}
\newcites{latex}{References}

\usepackage{lscape}

\usepackage{setspace}
\usepackage{tikz}
\usetikzlibrary{arrows}
\usepackage{multirow}

\def\om{\omega}
\usepackage{algorithm2e}
\usepackage{amsmath}
\usepackage{verbatim,color,amssymb}
\usepackage{amsmath}					
\usepackage{amsthm}					
\usepackage{natbib}
\usepackage{multirow}
\usepackage{setspace}
\usepackage[mathscr]{euscript}
\usepackage{fancyhdr}
\usepackage{enumitem}
\usepackage{graphicx}
\usepackage{subfigure}
\usepackage{lineno}
\usepackage{array,booktabs}
\usepackage{bm, bbm, bbold}
\usepackage{url}

\usepackage{lscape}

\usepackage{setspace}
\usepackage{tikz}
\usetikzlibrary{arrows}
\usepackage{multirow}

\newtheorem*{Proof*}{Proof}

\newtheorem{proposition}{Proposition}

\newtheorem{theorem}{Theorem}

\newtheorem{lemma}{Lemma}

\def\E{{\cal E}}

\def\N{{\mathrm N}}

\def\diag{\mathrm{diag}}
\def\Diag{\mathrm{Diag}}
\def\Ind{\mathbbm{1}}

\def\E{\mathrm{E}}

\def\tr{\mathrm{tr}}

\def\vect{\mathrm{vec}}
\def\vech{\mathrm{vech}}

\def\IG{\hbox{Inv-Ga}}

\def\Normal{\hbox{Normal}}

\def\Unif{\hbox{Unif}}

\def\bse{\begin{eqnarray*}}
\def\ese{\end{eqnarray*}}
\def\be{\begin{eqnarray}}
\def\ee{\end{eqnarray}}
\def\bq{\begin{equation}}
\def\eq{\end{equation}}

\def\trans{^{\rm T}}

\def\ba{\bm{a}}
\def\bA{\bm{A}}

\def\bB{\bm{B}}

\def\bC{\bm{C}}
\def\bd{\bm{d}}
\def\bD{\bm{D}}

\def\be{\bm{e}}
\def\bF{\bm{F}}

\def\bH{\bm{H}}

\def\bI{\bm{I}}
\def\bJ{\bm{J}}

\def\bK{\bm{K}}
\def\bL{\bm{L}}
\def\bM{\bm{M}}

\def\bP{\bm{P}}
\def\bq{\bm{q}}
\def\bQ{\bm{Q}}
\def\bR{\bm{R}}

\def\bS{\bm{S}}

\def\bT{\bm{T}}

\def\bU{\bm{U}}

\def\bV{\bm{V}}

\def\bW{\bm{W}}
\def\bx{\bm{x}}
\def\bX{\bm{X}}

\def\bY{\bm{Y}}

\def\bZ{\bm{Z}}
\def\bS{\bm{S}}

\DeclareMathOperator{\Eigen}{Eigen}

\newcommand{\bPhi}{\mbox{\boldmath $\Phi$}}

\newcommand{\btheta}{\mbox{\boldmath $\theta$}}
\newcommand{\bTheta}{\mbox{\boldmath $\Theta$}}

\newcommand{\bgamma}{\mbox{\boldmath $\gamma$}}
\newcommand{\bGamma}{\mbox{\boldmath $\Gamma$}}

\newcommand{\bSigma}{\bm{\Sigma}}

\newcommand{\bOmega}{\bm{\Omega}}

\newcommand{\bLambda}{\mbox{\boldmath $\Lambda$}}

\newcommand{\Frob}[1]{\Vert#1\Vert_{\mathrm{F}}}
\newcommand{\op}[1]{\Vert#1\Vert_{\mathrm{op}}}

\numberwithin{equation}{section}
\theoremstyle{plain}

\usepackage{bookmark}

\usepackage{float}
\usepackage{enumerate}
\usepackage{soul}
\usepackage{amsmath, amssymb} 
\usepackage{mathrsfs}
\usepackage{xcolor}

\usepackage[framemethod=tikz]{mdframed}

\usepackage{mhequ}
\usepackage{xcolor}
\usepackage{hyperref}

\usepackage{overpic}

\newtheorem{remark}{Remark}
\newtheorem{definition}{Definition}

\makeatletter

\newcounter{example}[section]

\usepackage{xcolor}

\newcommand{\bel}{\begin{equation}
        \begin{aligned}}

\newcommand{\eel}{ \end{aligned}
        \end{equation}
}

\usepackage{float}

\usepackage[noend]{algpseudocode}

\usepackage{subfiles}

\date{}

\begin{document}

\def\spacingset#1{\renewcommand{\baselinestretch}%
{#1}\small\normalsize} \spacingset{1}

\if1\blind
{
        \title{Bayesian Learning of Relational Graphs in Semiparametric High-dimensional Time Series}
\author{Arkaprava Roy$^1$, Anindya Roy$^2$ and Subhashis Ghosal$^3$\\ {\it $^1$University of Florida, Gainesville, USA,}\\ $^2${\it University of Maryland Baltimore County, Baltimore, USA,} \\ {\it and $^3$North Carolina State University, Raleigh, USA}
}} \fi

\if0\blind
{
  \title{Bayesian Learning of Relational Graphs in Semiparametric High-dimensional Time Series}
} \fi

\maketitle

\thispagestyle{empty} 
\begin{abstract}
We propose a novel approach for estimating the contemporaneous precision matrix for a high-dimensional time series. The generative model is based on several independent univariate time series, which are combined through orthogonal rotation and sparse linear transformation. A key feature of the approach is that any intrinsic relations among component time series, as specified by a graphical structure, are preserved across all time snapshots. We refer to the resulting process as Orthogonally-rotated Univariate Time series (OUT). Key structural properties of time series, such as stationarity and causality, can be easily accommodated in the OUT model. For Bayesian inference, we put suitable prior distributions. For inference, a pseudo-likelihood is constructed by using individual Whittle likelihoods for the univariate latent time series. An efficient Markov Chain Monte Carlo (MCMC) algorithm is developed for posterior computation. To establish posterior concentration for the model's parameters based on the Whittle likelihood-based pseudo-posterior, we develop a new general posterior convergence theorem by bounding moments of the likelihood ratio, and it may have a general appeal. We find that the known optimal posterior contraction rate for independent observations essentially prevails in the OUT model under very mild conditions on the temporal dependence, described in terms of the smoothness of the corresponding spectral densities. 
   Through a simulation study, we compare the accuracy of parameter estimation and graphical structure identification with other approaches. We apply the proposed methodology to analyze a dataset on different industrial components of the US gross domestic product between 2010 and 2019 and study the underlying graphical associations.
\end{abstract}

\newpage
\spacingset{1.9} 

\setcounter{page}{1}
\section{Introduction}
\label{sec:intro}

Understanding the intrinsic dependence structure among a set of random variables, assessed by the conditional dependence between two variables given the remaining variables, is a fundamental problem in statistics. Pairs of variables that are intrinsically dependent are connected by an edge, thus giving a graphical representation of the interrelations. 
Among probabilistic models for a relational graph, 
Gaussian Graphical Models (GGM) have the appealing property that conditional independence of a pair of variables given the others is equivalent to the zero-value of the corresponding off-diagonal entry of the precision matrix. A rich literature on estimating the graphical structure can be found in classical books such as \cite{Lauritzen1996, koller2009probabilistic}; see also \cite{Maathuis2019} and the references therein for more recent results.  The papers \cite{Jordan&Sejnowski:2001, meinshausen2006high, Yuan2007, friedman2008sparse, Buhlmann2011, wang2012bayesian, wang2012efficient, banerjee2014posterior,fan2016overview} provided some early developments of statistical estimation and structure learning methods in a GGM under both frequentist and Bayesian frameworks. Some of the follow-up developments have been on nonparanormal graphical model \citep{liu2012high, mulgrave2020bayesian}, pseudo-likelihood based CONCORD \citep{khare2015convex}, CLIME \citep{cai2011constrained}, Gaussian copula graphical model \citep{dobra2011bayesian,liu2012high}, other Markov random field based models \citep{wainwright2006high,yang2013poisson,chen2014selection,roy2020nonparametric}, etc. 

Since their introduction, GGM and closely related models have been applied across many disciplines, including neuroimaging, genetics, finance, political science, sociology, and network analysis. 
Linear correlation-based measures of association capture only marginal dependencies and can often be spurious.
However, marginal independence graphs may be relevant in some applications; our parameterization focuses on the conditional independence graph.
A typical modern application requires learning graphical structures in a high-dimensional setting. Most work on the inference for graphical models has been restricted to independently and identically distributed multivariate observations. 

In this paper, we use the contemporaneous precision matrix of a stationary time series to estimate the graph along the line of \cite{qiu2016joint,p:zha-17}. It turns out, though, that under the proposed model, the graphical structure encoded by the contemporaneous precision matrix encodes that of a stationary multivariate time series in the stronger sense as described in \cite{Dahlhaus2000}.  The multivariate data considered in this article are regularly spaced vector time series. However, we do not make modeling assumptions about the temporal dynamics using common parametric vector time-series models, treating the autocorrelation structure as a nuisance parameter in the graph learning problem. We propose a semi-parametric model, Orthogonally-rotated Univariate Time series (OUT), for vector time-series data. This model explicitly parameterizes the graphical structure, while the temporal structure is modeled nonparametrically. A Bayesian graph learning framework is used that provides efficient estimation and automatic uncertainty quantification through the resulting posterior distribution.

Several graphical structures may be of interest in a vector time-series model. A `contemporaneous stationary graphical' structure is where the graphical structure is encoded in the marginal precision matrix of a stationary Gaussian time series. In a contemporaneous stationary graph, the nodes are the coordinate variables of the vector time series. In a stationary time series, one may construct a graph with the nodes representing the entire coordinate processes. This would be a `stationary graphical' structure.  Conditional independence in such a graph can be expressed equivalently by the absence of partial spectral coherence between two nodal series; see \cite{Dahlhaus2000}. High-dimensional series estimation of such stationary graphs was investigated by \cite{jung2015graphical,fiecas2019spectral}. 
Some authors have studied time series in graphs and networks; see, for instance, \cite{basu2015network, ma2016joint}. Thus, our motivation is twofold: first, to develop a multivariate time-series framework for stationary data, and second, to characterize conditional-independence-based graphical associations among the component variables.

To model high-dimensional multivariate time-series data, a popular class of models exploiting lower-dimensional structures is the {\it Dynamic Factor Model} (DFM). The DFM finds use in many economic applications and other areas. Since its initial introduction to the econometric literature, a large body of work has emerged; see \cite{Forni2000TheGD, RBaiNg2002, Bai2003InferentialTF,carvalho2007dynamic, DEISTLER2010211, stock2012dynamic, Bottegal2015, Barogozzi2023}  and also \cite{lippi2022highdimensional} for a recent survey of related works along with other reduced-rank models such as \cite{matteson2011dynamic}. While the DFM is particularly well-suited to estimating latent lower-dimensional processes that drive the dynamics, the parameterization is unsuitable for estimating the graphical and stationary structures. The proposed semiparametric model in this paper closely resembles DFM, as both belong to the class of multivariate time-series models constructed from latent univariate time series. However, the proposed model allows a full-dimensional latent process without an idiosyncratic term. 
The direct parameterization of the contemporaneous precision matrix offers an advantage over traditional DFM for graph estimation, the main focus of this work. The proposed model and dynamic factor models both have symmetric autocovariance functions. 
The full-dimensional independent latent dynamic processes are analogous to the framework of {\it Independent Component Analysis}. 
Nevertheless, our primary objective is to consistently learn the graphical structure rather than to identify the latent source signal.
Existing methods for high-dimensional time series that estimate the graphical dependence through a sparse precision matrix, such as \cite{kolar2010estimating,chen2013covariance} may not maintain properties such as stationarity and causality in the estimated processes.

This article proposes a latent time series model based on a hidden low-dimensional structure within a high-dimensional time series. Specifically, let $\bY_t$ be a $p$-dimensional centered Gaussian time series observed at $T$ time points $\{1,\ldots,T\}$, modeled as $\bY_t=\bA \bZ_t$, where $\bZ_t$ is a $p$-variate centered time series consisting of independent components with unit variance at every $t$.
Then the dispersion matrix of $\bY_t$  is $\bSigma=\bA\bA\trans$ and the precision is $\bOmega=\bSigma^{-1}$. The low dimensionality in our model is hidden in $\bA$, a consequence of the requirement of sparsity in the marginal precision. When the $\bZ_t$  is Gaussian, the precision matrix, $\bOmega$, of $\bY_t$ determines the conditional independence structure in the time series at a given time point in that the $(j,k)$th off-diagonal entry of $\bOmega$ is $0$ if and only if $Y_{t,j}$ and $Y_{t,k}$ are independent given $\{Y_{l,t}: l=1,\ldots,p, \, l\ne j,k\}$. The necessity of the condition holds for non-Gaussian time series as well. However, the graph reduces to a partial correlation graph for non-Gaussian variables instead of a conditional independence graph. We let $\bSigma^{1/2}$ denote the positive definite Cholesky square root of $\bSigma$, and  write $\bA$ as $\bSigma^{1/2} \bU$, where $\bU$ is a $p\times p$ orthogonal matrix, leading to the representation $\bY_t=\bSigma^{1/2} \bU \bZ_t$ in terms of the uniquely identifiable parameter $\bSigma^{1/2}$ or the associated precision matrix $\bOmega$. This particularly allows for direct sparse parametrization of $\bOmega$. Since the model is represented by orthogonally rotating an ensemble of $p$ independent univariate time series, we shall call it the Orthogonally-rotated Univariate Time series (OUT) model. In the OUT model, the graphical dependence structure, encoded by the precision matrix elements, remains invariant over time, which will be our primary learning target. The OUT model shares some commonalities with \cite{chang2018principal}. However, our focus is on sparse estimation of the marginal precision, whereas that for \cite{chang2018principal} is performing principal component analysis. Given the different objectives, our model parameterization is tailored primarily to precision matrix estimation. Assuming a sparse structure of $\bOmega$, the graph may be identified by estimating $\bOmega$. We take a Bayesian route and induce a graphical structure by putting a prior on sparse precision matrices through a modified Cholesky decomposition and a sparse prior on the components of the Cholesky factor.  

In addition, the proposed OUT model provides a natural avenue for modeling stationary, causal, high-dimensional time series with an interpretable graphical dependence structure among components. This graph learning in high-dimensional stationary processes with a $p$-dimensional latent structure is possible because the latent time series are assumed to be independent, which facilitates the separation of the individual latent series during estimation. The spectral densities of the univariate time series are modeled using B-splines, with an additional assumption of low-rankness on the coefficients. Because of the underlying independence structure, the Whittle approximate likelihood can be conveniently used for inference in the OUT model. This allows a simpler expansion of the pseudo-likelihood for studying posterior convergence properties. 

Beyond the methodological innovations, we make two major theoretical contributions. 
\begin{itemize} 
\item [(a)] We develop a general posterior concentration theorem based on a misspecified likelihood (Theorem~\ref{thm:pseudoposterior general}), where the convenience of showing convergence under the simpler pseudo-likelihood can be transported to that under the true likelihood using a change of measure argument. Specifically, it relies on bounding moments of the likelihood ratio, and this technique may have a general appeal. 

\item [(b)] We establish that the posterior contraction rate (Theorems~\ref{thm:main} and \ref{thm:average Frobenius}) for the stationary precision matrix in terms of the Frobenius norm is given by $\sqrt{((p+s)\log p)/T}$ provided the spectral densities of the latent independent univariate time series are sufficiently smooth, where $s$ stands for the number of non-zero off-diagonal entries of the true precision matrix. The rate coincides with the optimal rate for the estimation of a sparse precision matrix with respect to the Frobenius norm for independent observations attained by the graphical lasso for independent data (cf. \cite{rothman2008sparse}) and also by an analogous Bayesian method developed by \cite{banerjee2015bayesian}. Thus, in the dependence setup of an OUT model, a precision matrix can be learned with the same level of accuracy as with independent observations under sparsity. Hence, the established rate cannot be improved further.
\end{itemize}


\section{Model and parameterization}
\label{sec:model}

\subsection{Notations and preliminaries}

We first describe the notations we follow in this article. Vectors are column-vectors by default and will be denoted by boldface lower-case English or Greek letters or upper-case English letters, while matrices will be denoted by boldface uppercase English or Greek letters. To denote a component of a vector or an entry of a matrix, we shall use the non-bold form of the letter along with the corresponding subscript(s). The transpose of a matrix $\bA$ is denoted by $\bA\trans$, and for a square matrix, $\tr(\bA)$, $\det(\bA)$, and $\bA^{-1}$ denote the trace, the determinant, and the inverse (if it exists) of $\bA$,  respectively. The vectorization $\vect(\bA)$ of a matrix $\bA$ stacks the columns of $\bA$ in a single column, while the half-vectorization $\vech(\bA)$ of a symmetric matrix is the same operation without repeating the identical entries originating from symmetry. We recall the following useful result. 
Finally, we use $\Eigen_j(\bA)$ to denote the $j$th eigenvalue of $\bA$.

We recall the following useful result. 
For a nonnegative definite matrix $\bA$, a square root $\bB$ is any matrix such that $\bB \bB\trans=\bA$, while the unique nonnegative definite square root is denoted by $\bA^{1/2}$ unless specified otherwise. The symbol $\bm{0}$ will stand for a zero-vector or a zero-matrix depending on the context, $\bm{1}$ a vector of ones, and $\bI_p$ will denote the $p\times p$-identity matrix. A diagonal matrix $\bA$ whose diagonal entries form a vector $\ba$ will be denoted by $\Diag(\bA)$, and conversely, we write $\ba=\diag(\bA)$. The same convention will also be used for block-diagonal matrices. The Kronecker product of two matrices will be denoted by $\otimes$ and the Hadamard product, standing for entrywise multiplication when the two matrices have identical dimensions, will be denoted by $\odot$. We shall use $\|\cdot\|_2$ to denote the Euclidean norm of a vector, $\|\cdot\|_\infty$ the maximum absolute value of the entries,  and $\|\bx\|_0$ for the number of non-zero entries of $\bx$. For a matrix $\bA$, the Frobenius norm $\Frob{\bA}=\|\vect(\bA)\|_2$ is the Euclidean norm of its vectorization,  the operator norm is given by $\op{\bA}=\sup\{ \|\bA \bx\|_2: \|\bx\|_2\le 1\}$ and the $\ell_0$-norm by $\|\bA\|_0=\|\vect(\bA)\|_0$.  Let $\Phi$ and $\phi$ stand for the cumulative distribution function and the probability density function of the standard normal, respectively.
The notation $\mathrm{N}_p({\mu}, \bSigma)$ stands for the $p$-variate normal measure with mean ${\mu}$ and dispersion matrix $\bSigma$, and its density will be denoted by $\phi_p(\cdot;{\mu}, \bSigma)$. We shall use the notations $\Ind$ for the indicator function and $\delta_0$ for the degenerate probability measure at $0$. Let $\E$ stand for the expectation operator and $\mathrm{D}$ for the dispersion matrix operator of a random vector. The symbols $\lesssim$, $\ll$, and $\asymp$ will respectively stand for domination by a constant multiple, strict domination in the order of growth or decay, and equality of the order of growth or decay for two given sequences of numbers. 

We also recall some useful relations between matrix norms. For a square matrix $\bA$, $\Frob{\bA}^2=\tr(\bA \bA\trans)$, $\op{\bA}\le \Frob{\bA}$, and $\Frob{\bA \bB}\le \min\{\Frob{\bA}\op{\bB}, \op{\bA}\Frob{\bB}\}$ for any two square matrices $\bA$ and $\bB$.

\subsection{The OUT model }

We begin with the formal introduction of an OUT model that is motivated to characterize $\bSigma^{1/2}$ in a way that allows sparse estimation of $\bOmega=\bSigma^{-1}$. 

\begin{definition}[OUT Process]
\label{def:OUT}
 A $p$-dimensional centered time series $\bY_t$, $t=1,\ldots, T$, is an Orthogonally-rotated Univariate Time series (OUT) with marginal positive definite covariance matrix $\bSigma$ if it admits a representation 
\begin{align}
\label{eq:OUT}
\bY_t= \bSigma^{1/2}\bU \bZ_t,
\end{align}
where $\bSigma^{1/2}$ is the lower triangular square root of $\bSigma$, $\bU$ is a $p\times p$ orthogonal matrix with determinant equal to $1$ and $\bZ_t=(Z_{1,t},Z_{2,t},\ldots,Z_{p,t})\trans$ is a vector of independent centered time-series with unit variances. 
\end{definition}

Since the distribution of any component of the process $(\bZ_t:t=1,\ldots,T)$ remains invariant under a sign change, we can assume without loss of generality that $\bU$ has determinant $+1$, by switching the sign of a component of $(\bZ_t:t=1,\ldots,T)$, if necessary. 
Thus, we can consider $\bU$ to be a special orthogonal matrix.

Recall that the autocovariance function of a centered multivariate time series $(\bX_t:t=0,1,\ldots)$ is defined by $\bGamma(h, t; \bX)=\E[\bX_t \bX_{t+h}\trans]$. 
A multivariate time series $(\bX_t:t=0,1,\ldots)$ is called (strictly) {\it stationary} if 
$(\bX_{t_1},\ldots,\bX_{t_k})$ and $(\bX_{t_1+h},\ldots,\bX_{t_k+h})$ 
have the same distribution 
$\forall\, k\ge 1,\; t_1,\ldots,t_k \in \mathbb{Z},\; h \in \mathbb{Z}.$
In that case, the {\it autocovariance function} (ACVF) denoted by $\Gamma(h; \bX)$  is free of $t$ and depends only on $h$. For a centered Gaussian time series, the converse also holds. A process is called {\it covariance stationary} if the ACVF is free of $t$. For a  covariance-stationary multivariate time series $(\bX_t: t=0,1,\ldots)$, the {\it spectral density matrix} $\bF(\om;\bX)$ is defined to be $(2\pi)^{-1}\sum_{h=0}^\infty \bGamma(h; \bX) e^{-i h \omega}$, $\om\in [-\pi,\pi]$. 
Finally, a time series $(\bX_t:t=0,1,\ldots)$ is called {\it causal} if for all $t$, $\bX_t$ can be expressed as ${H}_t({e}_t,{e}_{t-1},\ldots)$ for some process $({e}_t:t=0, \pm 1,\pm 2,\ldots)$, where ${H}_t$ is a vector-valued function possibly dependent on $t$. 

\begin{proposition}[Properties of OUT]
\label{prop:OUT properties}
The OUT process has the following properties. 
\begin{itemize}
\item [{\rm (i)}] The dispersion matrix of $\bY_t$ is given by $\mathrm{D}(\bY_t)=\bSigma$ and the precision matrix $\bOmega=\bSigma^{-1}$ for every $t$. 
    \item [{\rm (ii)}] The ACVF of the OUT process $\{\bY_t\}$ is $\bGamma(h, t; \bY) = \bSigma^{-1/2} \bGamma(h, t; \bZ)(\bSigma^{-1/2})\trans$, where $\bGamma(h, t; \bZ)$ is the ACVF  of $\{\bZ_t: t=0,1,\ldots).$ The function is symmetric in the lag-variable, i.e. $\bGamma(-h, t; \bY) = \bGamma(h, t; \bY)$.   
    \item [{\rm (iii)}] For every $t$, the eigenvectors of $\bGamma(h,t; \bY)$ are independent of $h$ but  the eigenvalues given by $\{ \gamma_j(h,t)=\E (Z_{j,t}Z_{j,t+h}): j=1,\ldots,p\}$ possibly depend on $h.$
    \item [{\rm (iv)}] If $(Z_{j,t}: t=1,2,\ldots)$, $j=1,\ldots,p$, are covariance-stationary with ACVF  $\gamma_j(h)$ and spectral densities $f_j(\omega)$, 
 then $(\bY_t: t=1,2,\ldots)$ is covariance-stationary with ACVF $\bGamma(h; \bY)$ and spectral density matrix $\bF(\omega; \bY)$. In addition, if all latent univariate time series are Gaussian, then the resulting OUT process is a strictly stationary Gaussian time series. 
  \item [{\rm (v)}]
    In the stationary case, the following relations hold
    \begin{eqnarray*}
       \bGamma(h; \bY) &=& \bSigma^{1/2}\bGamma(h; \bZ)(\bSigma^{1/2})\trans,\; \mbox{for all}\; h = 0, \pm1, \ldots \\
       \bF(\omega; \bY) &=& \bSigma^{1/2}\bF(\omega; \bZ) (\bSigma^{1/2})\trans, \; \mbox{for all}\; \omega \in [-\pi, \pi], 
    \end{eqnarray*}
    where $\bGamma(h; \bZ) = \Diag(\gamma_1(h), \ldots, \gamma_p(h))$ is the diagonal autocovariance matrix of $\bZ_t$ at lag $h$ with diagonal entries equal to the ACVF of the univariate components of $\bZ_t$ given at lag $h$, $\bF(\omega, \bZ) = \Diag(f_1(\omega), \ldots, f_p(\omega))$ is the diagonal spectral density matrix of $\bZ_t$ at frequency $\omega$ with diagonal entries equal to the univariate spectral densities of the components of $\bZ_t$ at frequency $\omega$. 
   \item [{\rm (vi)}] If the univariate time series $(Z_{j,t}: t=1,2,\ldots)$ is causal for each $j=1,\ldots,p$, then $(\bY_t: t=1,2,\ldots)$ is causal.  
\end{itemize}
\end{proposition}

These properties clarify both the structure and the practical implications of the OUT process. 
Property~(i) shows that the marginal dispersion and precision matrices of $\bY_t$ are time-invariant, which allows us to interpret conditional independence relationships through a fixed graphical structure. This property is essential for defining and estimating the stationary graph associated with the multivariate time series. 
Property~(ii) characterizes the autocovariance structure of $\bY_t$ in terms of the latent process $\bZ_t$, making explicit how temporal dependence propagates through the transformation; this representation is later used to derive the spectral density matrix. 
Property~(iii) shows that the eigenvectors of the autocovariance matrices do not depend on the lag, which simplifies both theoretical analysis and spectral decomposition by reducing the problem to univariate components. 
Property~(iv) establishes covariance stationarity of the OUT process under stationarity of the latent series making the analysis amenable to standard frequency-domain tools. 
Property~(v) provides explicit expressions for the autocovariance and spectral density matrices in the stationary setting, which are used in subsequent sections to specify the model and perform inference. 
Finally, Property~(vi) shows that the causal structure of the latent univariate processes carries over to the observed multivariate process, thereby justifying the use of causal representations and ensuring that the model admits a valid time-domain interpretation.

\subsection{Parameterization of the OUT model}

Given that we are considering a semi-parametric model for a high-dimensional time series, careful parameterization of the model's components is warranted, particularly one that permits a convenient lower-dimensional representation without compromising the model's flexibility. The following sections discuss the parameterization of the three components of the OUT model: the graphical parameters given by the identifiable square root of the $\bSigma$ matrix, the rotation matrix $\bU$, and the spectral densities of latent processes $(Z_{j,t}: t=0,1,\ldots)$, $ j = 1, \ldots, p$. 

\subsubsection{Parameterization of the latent processes}

Henceforth, we assume that all latent processes $(Z_{j,t}:t=1,2,\ldots)$, $j=1,\ldots,p$, are stationary. Since these processes are assumed to be independent and centered, under Gaussianity, their distributions are completely determined by their autocovariance $\gamma_j(h)$ or equivalently by their spectral densities $f_j(\omega)$. We assume Gaussianity of the univariate processes for likelihood development and parameterize them in terms of their spectral densities.

  For modeling the spectral densities, finite-dimensional parametric spectral density functions seem a natural choice. 
  However, we need $ \int_{-\pi}^\pi f_j(\omega) \cos(h \omega)d\omega=\gamma_j(0)=1$, as all $Z_{j,t}$ have unit variances. This constraint forces the  spectral densities $f_j$, $j=1,\ldots,p$, to be  symmetric probability densities on $[-\pi,\pi]$. Thus, if we choose a parametric model for spectral densities, the parameters of that model must satisfy the constraint of unit variance via the constraint on the spectral densities. 
  This makes widely used classes of univariate time series, such as autoregressive moving average (ARMA) processes, unsuitable for our purpose. For example, to make a univariate ARMA process with autoregressive order $o_1$, moving average order $o_2$, autoregressive parameters $a_1, \ldots, a_{o_1}$, moving average parameters $b_1, \ldots, b_{o_2}$,  innovation variance $\sigma_e^2$
  to have unit variance, severe non-linear restrictions have to be imposed on the parameters $a_1, \ldots, a_{o_1}, b_1, \ldots, b_{o_2}$ in addition to the complex constraints due to stationarity.  
  We consider a standard nonparametric technique of using a basis expansion with sufficiently many terms to specify the spectral densities, thereby reducing to a de facto parametric family. A B-spline basis is especially convenient because of its shape and order-preserving properties.  Writing $B_k^*=B_k/\int_0^1 B_k(u)du$ for the normalized B-splines with a basis consisting of $K$ many B-splines, we may consider a model indexed by the vector of spline coefficients $\btheta=(\theta_1,\ldots,\theta_K)$ given by 
\begin{align}
    \varphi(\omega;\btheta)= \sum_{k=1}^K \theta_{k}B_k^*(|\omega|/\pi),
    \quad 
    \theta_k\ge 0, \quad \sum_{k=1}^K \theta_k=1/2.    
    \label{eq:specden}
\end{align}

\begin{remark}[Separable covariance] 
\label{rem:separable}
A significant dimension reduction is possible by assuming all individual latent processes $Z_j$, $j=1,\ldots,p$, are identically distributed. In this case, the spectral densities are the same, and hence   $\btheta_1=\cdots=\btheta_p$. Letting $\bGamma_T$ stand for the common dispersion (correlation) matrix of $(Z_{j,1},\ldots,Z_{j,T})\trans$ for $j=1,\ldots,p$, the joint covariance matrix of $(Y_{1,1},Y_{2,1},\ldots, Y_{p-1,T}, Y_{p,T})\trans$ is given by $\bGamma_T\otimes\bSigma$. Thus, the occasionally used separable-covariance model, given by the Kronecker product of a correlation matrix and a covariance matrix, is a subclass of the proposed OUT model. 
\end{remark}

\subsubsection{Parameterization of the orthogonal rotation}

As mentioned earlier, for parameter identifiability, we can choose $\bU$ to be a special orthogonal matrix. We can thus use a suitable parameterization of the special orthogonal group. A popular choice is using the Cayley transformation
given by $\bU=(\bI_p-\bA)(\bI_p+\bA)^{-1}$, where $\bA$ is a skew-symmetric matrix. 

\subsubsection{Parameterization of graphical parameters}

Finally, we describe parameters that regulate the graphical structure. First, we make an interesting observation regarding the graphical structure of the time series in the OUT model. Define for any frequency $\omega \in [-\pi, \pi],$ $\bS(\omega) = \Diag(f_1(\omega), \ldots, f_p(\omega)).$
Then we have the following result.

\begin{proposition}
    Let $\bF_{\bY}(\omega)$ be the spectral density matrix of an OUT process $\bY_t$ at frequency $\omega$ and $\bF^{-1}_{\bY}(\omega) = \bB\bS^{-1}(\omega)\bB^T$ be the inverse spectral density matrix for $\bY_t$ where $\bB$ is a matrix square root of $\bOmega$. Then, $(\bF^{-1}_{\bY}(\omega))_{\ell,k}=0$ if and only if $(\bB\bB^T)_{\ell,k}=0$. 
    Thus, the contemporaneous precision matrix $\bOmega$ encodes the graphical structure of the time series in the strong sense: two different component processes $(Y_{i,t}: t=0,1,\ldots)$ and $(Y_{j,t}:t=0,1,\ldots)$ are independent conditionally on all the other $(p-2)$ coordinate processes if and only if the $(i,j)$th entry of $\bOmega$ is zero.
    \end{proposition}

\begin{proof}
    Write $\bS(\omega) =\Diag(s_{1},\ldots,s_{p})$. Then  
    \[(\bB\bS^{-1}(\omega)\bB^T)_{\ell,k}=({s_{k}s_{\ell}})^{-1}\sum_{a=1}^p b_{i,a}b_{i,k}\]
    and $(\bB\bB^T)_{\ell,k}=\sum_{a=1}^p b_{i,a}b_{i,k}$. Thus, $(\bB\bS^{-1}(\omega)\bB^T)_{\ell,k}=0$ if and only if $(\bB\bB^T)_{\ell,k}=0$. In our notation, $\bB\bB^T=\bOmega$, the marginal precision. Hence, the zero-entries of $\bOmega$ correspond to the zero-entries of the inverse-coherence at all frequencies. 
\end{proof}

The distribution of the OUT model is determined by $\bSigma$, $\bU$, and the distribution of the univariate series $(Z_{j,t}:t=1,2\ldots)$, $j=1,\ldots,p$. The number of free parameters in $\bSigma$ and that in $\bU$ are both $O(p^2)$ while $p$ functions control the distributions of the latent time series. Without any further sparsity assumption, the prior concentration rate near the true values would not be sufficiently fast to lead to a useful posterior contraction rate, and this would impose a strong growth restriction on $p$. However, a faster rate can be achieved under appropriate sparsity assumptions. It is sensible to assume that the off-diagonal entries of $\bOmega=\bSigma^{-1}$ are sparse, as this corresponds to a sparse conditional dependence structure among the component time series. For independent data of size $n$, such an assumption leads to the posterior contraction rate $\sqrt{((p+s)\log p)/n}$ for $\bOmega$ in terms of the Frobenius distance, where $s$ is the true number of non-zero sub-diagonal elements of $\bOmega$ signifying the number of edges connecting conditionally dependent variables. We shall show that the rate $\sqrt{((p+s)\log p)/T}$ for $\bOmega$ in terms of the Frobenius distance is also achievable assuming further structural restrictions on $\bU$ and $f_1,\ldots,f_p$.

We may consider the generalized square root $\bSigma^{1/2}=(\bI-\bL)^{-1}\bD^{-1}\bV$ of the covariance matrix, where the matrices $\bV$, $\bL$, and $\bD$ are respectively orthogonal, strictly lower triangular, and diagonal with positive entries. This leads to $\bZ_t=\bU\trans\bV\trans \bD(\bI-\bL)\bY_t$. However, $\bU$ and $\bV$ are not separately identifiable and $\bV$ is absorbed into the parameter $\bU$. Thus, we define the OUT process by $\bZ_t=\bU\trans \bD(\bI-\bL)\bY_t$. For clarity, this does not pose a problem since the specific components of the square root and the orthogonal matrices are not of interest. Our main focus is the marginal precision of $\bY_t$, which remains identifiable even if the two orthogonal matrices are combined into a single one. Also, it makes our final model order-invariant in the same spirit as in \cite{chan2018invariant, chan2024large,arias2024inference}. However, our overall model formulation differs and is specifically designed for sparse estimation of the marginal precision.

\section{Likelihood, Prior specification and posterior sampling}
\label{sec:prior&posterior}


To construct the likelihood for the OUT process, we use the inverse relation $\bZ_t = \bU\trans \bOmega^{1/2} \bY_t$ of \eqref{eq:OUT}, and use the independence of component series in $\bZ_t$. If Gaussianity is assumed, the joint distribution of $\bZ_t$, $t=1,\ldots,T$, can be written down in terms of the spectral densities, and hence the likelihood is expressible as a function of $\bgamma=(\bOmega,\bU,\btheta_1,\ldots,\btheta_p)$ upon the substitution of $\bZ_t$ in terms of $\bY_t$ and $(\bOmega,\bU)$. Even if these series are not Gaussian, this likelihood may be treated as a working likelihood. However, the joint distribution of $(Z_{j,t}:t=1,\ldots,T)$ involves a complicated matrix inversion of the autocorrelation matrix of $(Z_{j,t}: t=1,\ldots,T)$, making the actual likelihood very difficult to work with. A popular approach in time series is to use the  Whittle approximate likelihood for inference. Whittle's pseudo-likelihood is based on an approximate distributional result about the discrete Fourier transforms (DFT) of the time series data that asserts that at specific frequencies, the DFT values are approximately independent Gaussian variables with variances equal to half of the spectral density at these frequencies. Additional explanations on Whittle likelihood can be found in \cite{velasco2000whittle}. 

For a given $T$, define the set of Fourier frequencies as $\{\omega_l=2\pi(l-1)/T: l\in I_T\}$, where $I_T=\{1,\ldots,(T+1)/2\}$ for odd $T$ and $I_T=\{1,\ldots, T/2+1\}$ if $T$ is even. Let 
\begin{align}
W_{j,l,\mathrm{c}} & =(2\pi T)^{-1/2} \sum_{t=1}^T Z_{j,t} \cos (2 (t-1)\omega_l) , \quad l\in I_T, \label{eq:Whittle transform cos}\\ 
W_{j,l,\mathrm{s}} &=(2\pi T)^{-1/2} \sum_{t=1}^T Z_{j,t} \sin (2 (t-1)\omega_l), \quad l\in I_T, \, l\ne 1,T/2+1,\label{eq:Whittle transform sin}
\end{align}
$j=1,\ldots,p$, denote the cosine and sine transforms of the data at the Fourier frequencies. Then all these random variables are approximately independently distributed as $W_{j,l,\mathrm{c}}\sim \mathrm{N}(0,1/f_j(\omega_l))$,  $l\in I_T$, and $W_{j,l,\mathrm{s}}\sim \mathrm{N}(0,1/f_j(\omega_l))$, $l\in I_T$, $l\ne 1,T/2+1$. The assertion can be written as $\bW_{*j} \sim \mathrm{N}_{T} (0, \bLambda_j)$, where
\begin{eqnarray*}
 \bW_{j*} &=& (W_{j,1,\mathrm{c}}, W_{j,2,\mathrm{c}}, W_{j,2,\mathrm{s}},\ldots, W_{j,\nu,\mathrm{c}}, W_{j,\nu,\mathrm{s}})\trans, \nonumber  \\ 
\bLambda_{j,T} &=& \Diag(f_j(\omega_1),f_j(\omega_2),f_j(\omega_2),\ldots, f_j(\omega_\nu),f_j(\omega_\nu)), 
\end{eqnarray*}
for $T = 2\nu - 1$ and 
\begin{eqnarray*} 
\bW_{j*} &=(& W_{j,1,\mathrm{c}}, W_{j,2,\mathrm{c}}, W_{j,2,\mathrm{s}},\ldots, W_{j,\nu,\mathrm{c}}, W_{j,\nu,\mathrm{s}},W_{j,\nu+1,\mathrm{c}})\trans, \nonumber  \\ 
\bLambda_{j,T} &=& \Diag(f_j(\omega_1),f_j(\omega_2),f_j(\omega_2),\ldots, f_j(\omega_\nu),f_j(\omega_\nu), f_j(\omega_{\nu+1})),
\label{eq:whittle_trasf}
\end{eqnarray*}
for  $T = 2\nu$, where $\nu$ is a positive integere. 
Then the Whittle transform $\bW_{j*}$ represents the time series $(Z_{j,t}: t=1,\ldots,T)$ in the spectral domain, retaining the same information. The transformation works as an approximate decorrelator, leading to the simpler Whittle likelihood based on the approximate distributional assertion as an explicit function of the values of the spectral density at the Fourier frequencies. In our context, the series $(Z_{j,t}: t=1,\ldots,T)$, $j=1,\ldots,p$, are unobserved, so the likelihood has to be written in terms of the observed data $\bY$. To this end, define the corresponding Whittle transform of the series $(Y_{j,t}: t=1,\ldots,T)$ by replacing $(Z_{j,t}: t=1,\ldots,T)$ by $(Y_{j,t}: t=1,\ldots,T)$ in \eqref{eq:Whittle transform cos} and \eqref{eq:Whittle transform sin}. 
For $t=1,\ldots,T$, let $W_{jt}$ be the $t$th entry of $\bW_{j*}$, and $\bX_t=\bOmega^{-1/2} \bU \bW_{*t}$, where $\bW_{*t}$ stands for the $t$th column of $\bW= (\!( W_{jt})\!)$. We also write $\bX= \bOmega^{-1/2}  \bU \bW$. Then $\bX_t$, $t=1,\ldots,T$,  are approximately independent with $\bX_t\sim \mathrm{N}_{T} (0,\bOmega^{-1/2} \bU\bS_t \bU\trans \bOmega^{-1/2} )$,  
\begin{eqnarray}
\bS_1 &=& \Diag(f_1(\omega_1), \ldots, f_p(\omega_1)),\nonumber \\
\bS_{2k+1} = \bS_{2k} &=& \Diag(f_1(\omega_k), \ldots, f_p(\omega_k)), \;\; k = 1, \ldots, \lfloor(T-1)/2\rfloor, \nonumber \\
\bS_T &=& \Diag(f_1(\omega_{T/2}), \ldots, f_p(\omega_{T/2})), \; \mbox{for even } T.
\label{eq:st}
\end{eqnarray}

Hence, the log-likelihood based on Whittle's approximation is given by 
\begin{align}
\log q_T(\bgamma,\bX)=  -\frac{pT}{2}\log (2\pi)+ \frac12\sum_{t=1}^T \log\det \bM_t - \frac{1}{2} \sum_{t=1}^{T}  \bX_t\trans  \bM_t \bX_t,
   \label{eq:log-likelihood}
\end{align}
with 
$\bM_t=\bOmega^{1/2} \bU \bS_t^{-1} \bU\trans \bOmega^{1/2}$. Note that 
\begin{align}
\label{eq:det M}
\log\det \bM_t= \log \det \bOmega-\log \det \bS_t
= \log \det \bOmega-\sum_{j=1}^p \log f_j(\omega_{k(t)}),
\end{align}
where $k(t)=\lfloor t/2\rfloor +1$. 


In the following, we specify the prior distribution for each parameter of the vector $\bgamma$.  For some chosen constants $c_1,c_2,\sigma_d,\sigma_T,\sigma_\kappa,\lambda_L,\lambda_U>0$, the prior distributions of the individual parameters are given below.

$\bullet$ Prior for $\bOmega$: With $\bOmega$ expressed as $(\bI-\bL) \bD^2 (\bI-\bL)\trans + T^{-a}\bI_p$, where $a>1/2$ is fixed, independent priors are specified for $\bD=\diag(\bd)$ and $\bL$. Writing $\bC=(\bI-\bL)\bD$, an approximate $\bOmega^{1/2}$ can be set as $\bB=\bC + \frac{1}{2}T^{-a}\bC^{-\mathrm{T}}$. Specifically,
$\bB\bB^{T}=\bOmega+\frac{1}{4}T^{-2a}\bC^{-\mathrm{T}}\bC^{-1}$.

$-$ Prior for $\bd$: The components $d_1,\ldots,d_p$ of $\bd$ are independently distributed as inverse Gaussian distributions with density function 
$\pi_{d}(t) \propto t^{-3/2} e^{-(t-\xi)^2/(2t)}$, $t > 0$, for some $\xi>0$; see  
\citep{chhikara1988inverse}. This prior has an exponential-like tail near both zero and infinity. We put a weakly informative mean-zero normal prior with large variance $\sigma_d$ on $\xi$.  

$-$ Prior for $\bL$: Each sub-diagonal component of $\bL$ is independently distributed as $H_\lambda (Z)$ given $\lambda$, where $Z\sim \mathrm{N}(0,\sigma_T^2)$, $H_\lambda(z)=z \Ind\{ |z|>\lambda\}$ is the hard-thresholding operator at level $\lambda$ and $\lambda\sim \mathrm{Unif}[\lambda_L,\lambda_U]$.
This prior resembles a spike-and-slab prior, but is computationally advantageous. 

$\bullet$ Prior for $\bU$: Given that $\bU$ is multiplying the latent Gaussian series, we can fix the sign of the determinant of $\bU$ to be positive and hence choose $\bU$ as a special orthogonal matrix. The natural default prior is the Haar measure (i.e., the uniform distribution) on the Lie group of special $p\times p$ orthogonal matrices. However, the dimension of the space of orthogonal matrices is $p(p-1)/2=O(p^2)$. In high dimensions, the prior concentration rate will be weak and will slow down the target posterior contraction rate for $\bOmega$. Under the Cayley representation of the special orthogonal matrices, $\bU = (\bI_p - \bA)(\bI_p + \bA)^{-1}$ where $\bA$ is a skew-symmetric matrix. A better posterior contraction rate is possible if we impose a lower-dimensional structure on $\bA$. The skew-symmetric matrix $\bA$ has all diagonal entries zero, and sub-diagonal entries are unrestricted. The sparsity of $\bA$ can be ensured by specifying a spike-and-slab type prior for the sub-diagonal entries, where the slab is a fully supported distribution. The spike-and-slab type mechanism for sub-diagonal entries of $\bA$ is induced through a soft-thresholding operation using a thresholding parameter $\lambda_A$. We consider a uniform prior for $\lambda_A$.

$\bullet$ Prior for $\btheta_1,\ldots,\btheta_p$: If the spectral density is modeled parametrically like in an ARMA process, we can put a uniform distribution on $\bTheta = (\btheta_1,\ldots,\btheta_p)$. However, we do not make any parametric assumption and put a prior on a nonparametric class of functions as follows.  

If the spectral densities are modeled through the B-spline basis expansion \eqref {eq:specden}, then we put a prior through the parameterization of the splines. In this case, the ambient dimension becomes $pK\gg p$, where $K$ is the number of B-spline functions used in the expansion. The prior concentration for such a parameterization will be inadequate since $K$ will need to increase to infinity sufficiently fast to overcome the bias in the expansion. 

The problem can be overcome by only assuming a dimension reduction in the $\theta_{jk}$ values. A convenient representation that avoids the restriction of nonnegativity and the sum constraint is 
\begin{align} 
\theta_{jk}={\Psi(\kappa_{jk})}/\{2\sum_{l=1}^K\Psi(\kappa_{jl})\},  
\label{eq:spline in kappa}
\end{align} 
where $\Psi(u)=(1+u/(1+|u|))/2$ is a link function monotonically mapping the real line to the unit interval. Here, we used a bounded link function with only a polynomially small tail near negative infinity $\Psi(x)\sim 1/(2|x|)$ instead of a more popular exponential link. This enables more efficient bounding of the coefficients required to achieve the optimal posterior contraction rate. 
We note that this representation is not unique due to the sum constraints. 

A dimension reduction to the order $p$ can be achieved through a low-rank decomposition: 
\begin{align} 
\kappa_{jk}=\sum_{r=1}^R\xi_{jr}\eta_{kr}, \quad j=1,\ldots,p, \; k=1,\ldots,K, 
\label{eq:kappa low rank tensor}
\end{align}
in which case, the dimension reduces to $R(p+K)$. Since $K$ can typically be considered to be a low power of $T$, assuming enough smoothness in the spectral densities, while $p$ is typically a high power, the order will be $O(p)$, assuming that $R$ can be chosen fixed. This will suffice for our rate calculations. We note that redundancy in the representation is also eliminated through the assumed dimension reduction. Moreover, it contains the separable case of equal spectral densities when $\xi_j=1$ for all $j$. The alternative representation using sums $\kappa_{jk}=\xi_j +\eta_k$ covers only the separable case due to the cancellation effect of a constant for each $j$. 

Finally, since an appropriate value of the rank $R$ is unknown, we consider an indirect automatic selection of $R$ through a cumulative shrinkage prior \citep{bhattacharya2011sparse} on $\xi_{jr}$ $j=1,\ldots,p$, $r=1,\ldots,R$: For all $r=1,\ldots,R$, we put independent priors 
\begin{align*} 
\xi_{jr}|v_{jr},\tau_{r}&\sim \mathrm{N}(0,v_{jr}^{-1}\tau_{r}^{-1}), \quad v_{jr}\sim \mathrm{Gamma}(\nu_1,\nu_1),\quad \tau_{r}=\prod_{t=1}^r\Delta_{i},\\
\eta_r=(\eta_{1r},\ldots,\eta_{Kr}) &\sim\mathrm{N}_K (0,\sigma_{\kappa}\bP^{-1}), \quad  
\sigma_\xi,\sigma_{\kappa}\sim\IG(c_1,c_1),
\end{align*} 
where $\Delta_{1}\sim \mathrm{Gamma}(\kappa_{1}, 1)$ and $\Delta_{i}\sim \mathrm{Gamma}(\kappa_{2}, 1)$, $i\geq 2$. The parameters $v_{jr}$, $j=1,\ldots,p$, $r=1,2,\ldots$, control the local shrinkage of the elements in $\xi_{jr}$, whereas $\tau_r$ controls the shrinkage of the $r$th column.
The matrix $\bP$ is the second-order difference matrix to impose smoothness. Specifically, $\bP=\bQ\trans \bQ$, where $\bQ$ is a $K\times(K+2)$ matrix such that $\bQ\eta_r$ computes the second order differences in $\eta_r$.
The above prior thus induces smoothness in the coefficients because it penalizes $\sum_{j=1}^{K}(\Delta^2 \eta_r)^2 = \eta_r\trans\bP\eta_r$, the sum of
squares of the second-order differences in $\eta_r$ (see \cite{eilers1996flexible}).
On $K$, we put a prior with a Poisson-like tail $e^{-K \log K}$. 

Section S1 of the supplementary materials includes posterior sampling steps. Most parameters are sampled using the gradient-based Langevin Monte Carlo (LMC) and the rest by the adaptive Metropolis--Hastings algorithm \citep{haario2001adaptive}. In the supplementary materials, we also provide details on MCMC initialization.

\section{Convergence of the posterior distribution}
\label{sec:convergence}

In this section, we obtain the contraction rate $\epsilon_T$ of the posterior distribution of $\bgamma$ near its true value $\bgamma_0$, that is we identify a sequence $\epsilon_T$ satisfying  
$\Pi (\| \bgamma-\bgamma_0\|_2 >m_T \epsilon_T|\bY^{(T)})\to 0$ in probability under the true distribution for every sequence $m_T\to \infty$ as $T\to\infty$, where $\bY^{(T)}=\{\bY_t\}_{t=1}^T$. The general theory of posterior contraction rate, developed in \cite{ghosal2000convergence,ghosal2007convergence} and synthesized in \cite{ghosal2017fundamentals}, is hard to apply directly in the present context since the likelihood used to define the posterior distribution is based on the Whittle approximation and hence gives only a pseudo-posterior distribution. A possible route could be using the mutual contiguity of the Whittle measure with respect to the true measure \cite{choudhuri2004contiguity,choudhuri2004bayesian}. While the contiguity holds for the distribution of each latent process individually, the joint distributions of these fail to be contiguous due to the increasing dimension.  
Some extensions are offered by the contraction rate theory for misspecified models in \cite{kleijn2006misspecification}, and the theory of posterior contraction to pseudo-posterior distribution in high-dimensional parametric models provided by \cite{atchade2017contraction}. The former requires identically distributed observations, the computationally demanding calculation of the Kullback-Leibler projection of the true density, and the covering number for testing. Then it yields a rate in terms of the Hellinger distance only, which does not serve our purpose. Atchad\'e's theory uses an expansion of the log-pseudo-likelihood function to construct tests. While it is a promising approach for the present problem, obtaining appropriate bounds for the remainder of the expression globally in each split-cone, adapted to the dimension of that cone, is a daunting task. 

Thus, we first develop a new general posterior concentration result for studying pseudo-posterior (Thoerem~\ref{thm:pseudoposterior general}). 
Using that, we can directly use the properties of the Whittle pseudo-likelihood to obtain a general posterior contraction theorem for general observations (a variant of Theorem~8.19 of \cite{ghosal2017fundamentals}) with simplifications and minor adjustments.

To show posterior convergence, we define the likelihood equivalently in a slightly different form.  For a given value of the parameters $\btheta = (\bSigma, \bU, f_1, \ldots, f_p)$, the Whittle transformed data $\bW = (\bW_{1*}\trans, \ldots, \bW_{p*}\trans)\trans = \bU\trans \bSigma^{-1/2}\bY \bF = \bZ\bF$, as defined in \eqref{eq:whittle_trasf}, has the true and Whittle approximate densities given by 
\begin{align}
  p_{\btheta}^{(T)} = \prod_{j=1}^p \phi_T(\bW_{j*};\bm{0}, \bF\trans\bGamma_{j,T}\bF),  \quad 
  q_{\btheta}^{(T)} = \prod_{j=1}^p \phi_T(\bW_{j*};\bm{0}, \bLambda_{j,T}), 
  \label{eq:lkhds}
\end{align}
respectively, where $\bGamma_{j,T}$ is the variance of $\bZ_{j} = (z_{1,j}, \ldots, z_{T,j})\trans.$ 
For a given ${c}=(c_{1j}, c_{2j}: j=1,\ldots, p)$, consider the parameter space 
\begin{align}
\Theta_{{c}} = \{ (\bOmega, \bU, f_1, \ldots, f_p): \bOmega \in \mathcal{S}_T^{++}, \,\bU \in \mathrm{SO}(T),\, \int f_j = 1,\, 0 < c_{1j} < f_j < c_{2j} < \infty \},
\label{eq:parm_space}
\end{align}
where $\mathcal{S}_T^{++}$ is the cone of $T\times T$ positive definite matrices and $\mathrm{SO}(T)$ is the group of $T\times T$ special orthogonal matrices. 

\begin{theorem}[General pseudo-posterior contraction rate]
\label{thm:pseudoposterior general}
Let $q_\theta^{(T)}$ be a statistical models for a sequence of observations $\bY^{(T)}$ indexed by $\theta\in \Theta_T$, $T=1,2,\ldots$, and let $p_0^{(T)}$ be the true distribution of $\bY^{(T)}$. Assume that for some fixed $b>1$ and $\theta_0=\theta_{0,T}\in \Theta_T$,   
\begin{align}
    \label{eq:moment control}
   \int (p_0^{(T)}/q_{\theta_0}^{(T)})^b q_{\theta_0}^{(T)} \lesssim G_T<\infty
\end{align}
for some sequence $G_T>0$. 
Let $d_T$ be a semimetric on $\theta_T$. Let $\Pi$ stand for a prior distribution for $\theta\in \Theta_T$. Suppose that there exist subsets a sequence $\epsilon_T\to 0$, $T\epsilon_T^2\to \infty$, sets $\Theta_T^*\subset \Theta_T$, a covering $\{\Theta_{T,j}^*: j=1,\ldots,N_T\}$ of $\Theta_T^*$, test functions $\phi_{T,j}$ such that the following conditions hold.
\begin{enumerate}
    \item [{\rm (i)}] $-\log \Pi ( B(\theta_0,\epsilon_T)) \lesssim T\epsilon_T^2$, where $B(\theta_0,\epsilon_T)= \{\theta: \int p_{0}^{(T)} \log (q_{\theta_0}^{(T)}/q_\theta^{(T)}) \le T\epsilon_T^2 \}$; 
     \item [{\rm (ii)}] $\int \phi_{T,j} q_{\theta_0}^{(T)} \le e^{-K_0 T\epsilon_T^2}$ and $\sup \{ \int (1-\phi_{T,j}) q_\theta^{(T)}: d_T(\theta_0,\theta)>\epsilon_T, \theta\in \Theta_{T,j}^*\} \le e^{-K_0 T\epsilon_T^2}$ for some $K_0>0$;
     \item [{\rm (iii)}]  $\log N_T \lesssim T\epsilon_T^2$;
     \item [{\rm (iv)}] $e^{cT\epsilon_T^2} \Pi (\Theta_T\setminus \Theta_T^*)/  \Pi (B(\theta_0,\epsilon_T))\to 0$ for any fixed $c>0$;
     \item [{\rm (v)}] $\log G_T \lesssim T\epsilon_T^2$.
\end{enumerate}
Then for every $m_T\to \infty$, $\Pi (d_T( \theta_0,\theta)> m_T \epsilon_T|\bY^{(T)})\to 0$ in $p_{0}^{(T)}$-probability. 
\end{theorem}

Theorem~\ref{thm:pseudoposterior general} could be of independent interest in deriving the posterior contraction rate of a pseudo-posterior from that of a standard posterior convergence theorem in a well-specified model. 
If $G_T$ is a bounded sequence, then condition (i) can be easily verified from a standard condition on prior concentration in Kullback-Leibler type neighborhoods in the working model $\{q_\theta^{(T)}: \theta\in \Theta_T\}$ because then by H\"older's inequality, 
\begin{align}
    \label{eq:KL through Holder}
 \int p_{\theta_0}^{(T)}\log (q_\theta^{(T)}/q_{\theta_0}^{(T)})\le \big(\int q_{\theta_0}^{(T)}|\log  (q_\theta^{(T)}/q_{\theta_0}^{(T)})|^{b/(b-1)} \big)^{(b-1)/b} G_T^{1/b}.
 \end{align}
The conditions (ii), (iii), and (iv) are commonly assumed in a posterior contraction rate theorem for the model $\{q_\theta^{(T)}: \theta\in \Theta_T\}$ under the true distribution $q_{\theta_0}^{(T)}$.  
However, in our situation, the constant $G_T$ is exponential in $p$, which grows rapidly due to the high dimensionality of $p$, and hence the estimate in  \eqref{eq:KL through Holder} is not useful. However, using the normality of $p_\theta^{(T)}$ and $q_\theta^{(T)}$, we shall obtain a useful bound and derive the posterior contraction rate in Theorem~\ref{thm:main} below. 

Another important technical tool is the construction of tests to establish posterior contraction rates, applying the general theory of \cite{ghosal2000convergence} along the lines of the earlier posterior consistency theory of \cite{schwartz1965bayes}. Here, instead of a canonical metric like the average squared Hellinger distance for independent observations that automatically leads to exponentially powerful tests, we use the R\'eyni divergence or the negative log-affinity in the Whittle model and use the likelihood ratio test to construct exponentially powerful tests. This is because converting the average squared Hellinger distance to the Frobenius distance on $\bOmega$ requires boundedness of the parameter space. But as it was observed earlier by \cite{ning2020bayesian,jeong2021unified,shi2021bayesian}, the use of the slightly stronger metric avoids this problem. However, the test has to be constructed directly, as the general Le Cam-Birg\'e tests are not available for this distance. As in the cited papers, we use the Neyman-Pearson likelihood ratio test for the true null against a distant alternative and show that the same test still has exponentially small error probabilities at parameters near the alternative by controlling a moment of the density ratio. 
 
We make the following assumptions about the true parameters. 

\begin{description}
\item [(A1)] The true precision matrix $\bOmega_0$ has eigenvalues that are bounded and bounded away from zero and has $s$ non-zero sub-diagonal entries. 
\item [(A2)] The true orthogonal transformation matrix $\bU_0$ has Cayley representation $(\bI_p-\bA_0)(\bI_p+\bA_0)^{-1}$, where $\bA_0$ has at most $s$ non-zero sub-diagonal entries, all lying in a fixed bounded interval. 
\item [(A3)] The true spectral densities $f_{0,1},\ldots,f_{0,p}$ of the latent processes are positive and H\"older continuous of smoothness index $\alpha$ for some $\alpha>0$ (cf.,  Definition~C.4 of \cite{ghosal2017fundamentals}) such that for some uniformly bounded sequence $\theta_{jk}^*$ of the form \eqref{eq:spline in kappa} and \eqref{eq:kappa low rank tensor}, it holds that 
\begin{align}
\label{eq:spline approximation}
    \max_{1\le j\le p} \{ \sup \{|f_{0,j}(\omega)-\sum_{k=1}^K \theta_{jk}^* B_k^*(\omega)|: \omega \in [0,1]\}\lesssim K^{-\alpha},
\end{align}
that is, the approximation ability of the splines for the true spectral densities is not affected if the coefficients are restricted by the dimension-reduction condition. 
\item [(A4)] 
Growth of dimension: $\log p\lesssim \log T$. 
\end{description}

Let $\bM_{t0}$ stand for the true value of $\bM_t$, $t=1,\ldots,T$. We consider a prior as described in Section~\ref{sec:prior&posterior}. For the rate below, $a>1/2$ implies $\sqrt{p}\,T^{-a}=o(\epsilon_T)$, so the nugget is asymptotically negligible.
Then we derive the following contraction rate for the posterior distribution $\Pi$ given the data based on the Whittle likelihood.  
We apply Theorem~\ref{thm:pseudoposterior general} and thus need to establish the moment condition in \eqref{eq:moment control}. The following result concerns that. 

\begin{lemma}
\label{lem:contiguity}
Let  $p_{\btheta}^{(T)}$ and $q_{\btheta}^{(T)}$ be the likelihoods based on the true Gaussian measure and Whittle's approximate measure. Let $c_{ij}<c_{2j}$ be such that for some fixed $b>1$, $\min\{c_{2j}/(c_{2j}-c_{1j}): j=1,\ldots,p\}>b$. 
Then uniformly for any $\btheta \in \bTheta_{{c}}$, 
\begin{align}
\log \int (p_\theta^{(T)}/q_\theta^{(T)})^b q_\theta^{(T)}\asymp p.
\end{align}
\end{lemma}

The proof is in the Supplementary Section S2, and it relies on the results from \cite{GrenanderSzego1984} for eigenvalues of the covariance matrix relative to the spectral density. Now, we have our first posterior contraction rate result.

\begin{theorem}
\label{thm:main}    
For $\epsilon_T=\max\{\sqrt{(p+s)/T}, T^{-\alpha/(2\alpha+1)}\} \sqrt{\log T}$ and any $m_T\to \infty$, for some constant $c_0>0$ that may be chosen as large as we please, we have that 
\begin{enumerate}
\item [{\rm (1)}] $\Pi( T^{-1} \sum_{t=1}^T \sum_{j=1}^p\min\{\Eigen^2_j(\bM_{t0}^{-1/2}(\bM_t -\bM_{t0})\bM_{t0}^{-1/2}), c_0\} \ge m_T^2 \epsilon_T^2|\bY^{(T)})\to 0$;  
    \item [{\rm (2)}]  $\Pi( T^{-1} \sum_{t=1}^T \sum_{j=1}^p\min \{\Eigen_j^2(\bM_{t0}^{1/2}(\bM_t^{-1}-\bM_{t0}^{-1})\bM_{t0}^{1/2}), c_0\} \ge m_T^2 \epsilon_T^2|\bY^{(T)})\to 0$,
    \end{enumerate}
in probability under the true distribution.
\end{theorem}

The conclusions in the theorem are somewhat indirect, in that they involve the average of the squared eigenvalues of the normalized differences of covariance or precision matrices for the Fourier transforms of the component series. If the diagonal entries in the Cholesky decomposition and the spectral densities are known to lie within a fixed compact subinterval of the positive real line, then the operation of computing the minimum is superfluous. Subsequently, the corresponding convergence yields rates in terms of the Frobenius norms of the stationary covariance and precision matrices of the multivariate time series. 

\begin{theorem}
\label{thm:average Frobenius}
Consider the setup of Theorem~\ref{thm:main}. If the diagonal entries $d_1,\ldots,d_p$ and the range of the functions $f_1,\ldots,f_p$ lie in a fixed compact subinterval of $(0,\infty)$, and the corresponding priors are replaced by the ones with domains truncated to compact intervals sufficiently large to contain the true values in the interior, then 
the following assertions hold: 
\begin{enumerate}
 \item [{\rm (a)}]  $\Pi( T^{-1} \sum_{t=1}^T \Frob{\bM_t-\bM_{t0}}^2\ge m_T^2 \epsilon_T^2|\bY^{(T)})\to 0$;
  \item [{\rm (b)}] 
 $\Pi( T^{-1} \sum_{t=1}^T \Frob{\bM_t^{-1}-\bM_{t0}^{-1}}^2 \ge m_T^2 \epsilon_T^2|\bY^{(T)})\to 0$; 
    \item [{\rm (c)}] $\Pi(\Frob{\bSigma-\bSigma_0} \ge m_T \epsilon_T|\bY^{(T)})\to 0$;
      \item [{\rm (d)}] $\Pi(\Frob{\bOmega-\bOmega_0}\ge m_T \epsilon_T|\bY^{(T)})\to 0$,
    \end{enumerate}
in probability under the true distribution.
\end{theorem}

If the spectral densities are sufficiently smooth so that $T^{1/(1+2\alpha)}$ is dominated by $p+s$, then the rate $\epsilon_T=\sqrt{(p+s)/T)\log T}$ prevails. Consequently, Parts (c) and (d) of the theorem conclude that the posterior contraction rate for the covariance and precision matrix in the OUT model coincides with the known optimal rate for independent and identically distributed observations, provided the stated boundedness conditions hold. The condition of high smoothness of spectral densities is tantamount to less temporal dependence of the time series. For instance, ARMA-type processes have infinitely smooth spectral densities and satisfy the requirement. 

The proof of Theorem~\ref{thm:main}  consists of verifying the conditions of Theorem~\ref{thm:pseudoposterior general} and is somewhat long. It is convenient to break the proof into auxiliary lemmas that verify individual conditions, as presented below.

\begin{lemma}
\label{lem:multinormal estimates}
Let $\bOmega_1,\bOmega_2$ be positive definite matrices.  Then for some constant $c_1>0$ and $c_2>0$, 
\begin{align} 
\int \sqrt{\phi_p (\bx;\bm{0},\bOmega_1^{-1})\phi_p (\bx;\bm{0},\bOmega_2^{-1})} d\bx & = \frac{(\det \bOmega_1)^{1/4} (\det \bOmega_2)^{1/4}}{(\det ((\bOmega_1+\bOmega_2)/2))^{1/2}} \notag \\
& \le \exp[-c_2\sum_{j=1}^{p}\min\{\Eigen^2_j(\bOmega_1^{-1/2}(\bOmega_1-\bOmega_2)\bOmega_1^{-1/2}),c_1\}].
\label{eq:affinity bound}
\end{align}
If further, $2\bOmega_2-\bOmega_1$ is positive definite and $\Frob{\bOmega_2-\bOmega_1}\le \op{\bOmega_1^{-1}}/3$, then 
\begin{align} 
\int \big( \frac{\phi_p (\bx;\bm{0},\bOmega_2^{-1})}{ \phi_p (\bx;\bm{0},\bOmega_1^{-1})}\big)^2 d\mathrm{N} (\bm{0},\bOmega_1^{-1}) & =\frac{\det \bOmega_2}{\sqrt{\det \bOmega_1} \sqrt{\det (2\bOmega_2-\bOmega_1)}}\notag \\
&\le \exp\{15 \op{\bOmega_1^{-1}}^2\Frob{\bOmega_2-\bOmega_1}^2 \}.
\label{eq:LR second moment bound}
\end{align}

Further, 
\begin{align} 
\int \sqrt{\phi_p (\bx;\bm{0},\bOmega_1^{-1})\phi_p (\bx;\bm{0},\bOmega_2^{-1})} d\bx\le \exp[-c_2\sum_{j=1}^{p}\min\{\Eigen^2_j(\bOmega_1^{1/2}(\bOmega_1^{-1}-\bOmega_2^{-1})\bOmega_1^{1/2}),c_1\}]
\label{eq:affinity bound2}
\end{align}
and if $2\bOmega_1^{-1}-\bOmega_2^{-1}$ is positive definite and $\Frob{\bOmega_2^{-1}-\bOmega_1^{-1}}\le \op{\bOmega_1}/3$, then 
\begin{align} 
\int \big( \frac{\phi_p (\bx;\bm{0},\bOmega_2^{-1})}{ \phi_p (\bx;\bm{0},\bOmega_1^{-1})}\big)^2 d\mathrm{N} (\bm{0},\bOmega_1^{-1}) \le \exp\{15 \op{\bOmega_1}^2\Frob{\bOmega_2^{-1}-\bOmega_1^{-1}}^2 \}.
\label{eq:LR second moment bound2}
\end{align}

In both \eqref{eq:affinity bound} and \eqref{eq:affinity bound2}, the constant $c_1$ may be chosen as large as we please at the expense of making $c_2$ smaller.
\end{lemma}

\begin{lemma}
\label{lem:eigenvalues}
The eigenvalues of $\bM_{t0}$ lie uniformly in some compact interval $[b_1,b_2]\subset (0,\infty)$. 
\end{lemma}

\begin{lemma}
\label{lem:U bound}
Let $\bA_1,\bA_2$ be $(p\times p)$-skew-symmetric matrices such that $\|\bA_1\|_\infty, \|\bA_2\|_\infty \le B$ for some $B>0$. Then for $\bU_1=(\bI_p-\bA_1)(\bI_p+\bA_1)^{-1}$, $\bU_2 =(\bI_p-\bA_2)(\bI_p+\bA_2)^{-1}$, we have that $\Frob{\bU_1-\bU_2}\le (1+\sqrt{p+p(p-1)B^2}) \Frob{\bA_1-\bA_2}$. 
\end{lemma}

\begin{lemma}
\label{lem:Omega bound}
Let $\bL_1,\bL_2$ be $(p\times p)$-strictly lower-triangular matrices and $\bD_1,\bD_2$ be diagonal matrices with entries in $(0,B)$ for some $B>0$ and $\op{\bI-\bL_k}\le C$ for $k=1,2$ and some $C>0$. Then for $\bOmega_k=(\bI_p-\bL_k)\bD^2_k (\bI_p-\bL_k)\trans+T^{-a}\bI_p$, $k=1,2$, we have that $\Frob{\bOmega_1-\bOmega_2}\le 2B^2 \Frob{\bL_1-\bL_2}+ C\Frob{\bD_1^2-\bD^2_2}$. 
\end{lemma}

\begin{lemma}
\label{lem:prior concentration}
For $\epsilon_T$ as in Theorem~\ref{thm:main}, $-\log \Pi ( \max_{1\le t\le T} \Frob{\bM_t-\bM_{t0}}^2 \le \epsilon_T^2) \lesssim T \epsilon_T^2$. 
\end{lemma}

\begin{lemma}
\label{lem:test}
Let $\bM_{t1}$, $t=1,\ldots,p$, be positive definite matrices satisfying the condition that $T^{-1}\sum_{t=1}^T  \sum_{j=1}^p\min\{\Eigen^2_j(\bM_{t0}^{-1/2}(\bM_t -\bM_{t0})\bM_{t0}^{-1/2}), c_0\}>\epsilon$ for some $c_0>0$ and $\epsilon>0$. Then there exists a test function $\phi_T$ such that for any positive definite matrices $\bM_{t2}$, $t=1,\ldots,T$    with $\max\{\Frob{\bM_{t2}-\bM_{t1}}: 1\le t\le T \}< \epsilon /(C\op{_{1t}^{-1}} )$ for some sufficiently large constant $C>0$, we have that  
$\int \phi_T \prod_{t=1}^T d\mathrm{N}_p(\bm{0}, \bM_{t0}^{-1}) \le e^{-K_0 T\epsilon^2}$ and $ \int (1-\phi_T)  d\mathrm{N}_p(\bm{0}, \bM_{t2}^{-1}) \le e^{-K_0 T\epsilon^2}$ for some constant $K_0>0$.

The same conclusion holds if instead the condition $\max\{\Frob{\bM_{t2}^{-1}-\bM_{t1}^{-1}}: 1\le t\le T \}< \epsilon /(C\op{\bM_{1t}^{-1}} )$ for some sufficiently large constant $C>0$. 

Moreover, the constant $c_0$ may be chosen as large as we please at the expense of making $K_0$ smaller. 
\end{lemma}

\begin{lemma}
\label{lem:covering}
For a constant $L>0$, put $s_T=L T\epsilon_T^2/\log p$ and $b_T=L\sqrt{T}\epsilon_T$. Define the sieve $\mathcal{M}_T$ as the collection of $(\bM_t = \bOmega^{1/2} \bU \bS_t^{-1} \bU\trans \bOmega^{1/2}:1\le t\le T)$ satisfying
\begin{enumerate}
    \item [{\rm (i)}] $\bOmega=(\bI-\bL) \bD^2 (\bI-\bL)\trans+T^{-a}\bI_p$, $\|\bL\|_0\le s_T$, $\|\bL\|_\infty \le b_T$, $\bD=\Diag(d_1,\ldots,d_p)$, $\max(d_j, d_j^{-1}:1\le j\le p) \le b_T$;
    \item [{\rm (ii)}] $\bU=(\bI_p -\bA) (\bI_p+\bA)^{-1}$,  $\|\bA\|_0\le L {T}  \epsilon_T^2/\log p$, $\|\bA\|_\infty \le L \sqrt{T} \epsilon_T$; 
    \item [{\rm (iii)}] $\bS_t$, $t=1,\ldots,T$, given by \eqref{eq:st}, \eqref{eq:spline in kappa} and \eqref{eq:kappa low rank tensor} such that $\max\{|\xi_{jr}|: 1\le j\le p, 1\le r \le R\} \le \sqrt{T}\epsilon_T$, $\max\{|\eta_{kr}|: 1\le k\le K, 1\le r \le R\} \le L \sqrt{T}\epsilon_T$, $K\le L T\epsilon_T^2/\log T$.
\end{enumerate}

On this sieve, the nugget and the bounds in item~{\rm (i)} give
\begin{align}
\op{\bOmega^{-1}}&\le T^a,\label{eq:sieve Omega inverse bound}\\
\op{\bOmega}
&\le \Frob{\bOmega}
\le \sqrt{p}\big\{(1+\sqrt{s_T}\,b_T)^2b_T^2+T^{-a}\big\}.
\label{eq:sieve Omega bound}
\end{align}
Indeed, $\bOmega\succeq T^{-a}\bI_p$ and $\op{\bI-\bL}\le 1+\Frob{\bL}\le 1+\sqrt{s_T}\,b_T$. Thus, both bounds are polynomial in $T$ under (A4).

Then $\Pi ((\bM_t:1\le t\le T)\not\in \mathcal{M}_T)\le e^{-L' T \epsilon_T^2}$ for some $L'>0$ that can be chosen as large as we please by making $L>0$ sufficiently large, and $\mathcal{M}_T$ can be split into $N$ pieces $\mathcal{M}_{T,l}$, $l=1,\ldots,N$, $\log N\lesssim T \epsilon_T^2$, such that there exist test functions 
$\phi_{T,l}$, $k=1,\ldots,N$, satisfying the conditions that for some $K_0>0$, $m>0$, for all $(M_t: 1\le t\le T)\in \mathcal{M}_{T,l}$,
$$\int \phi_{T,l} d\N_p(\bm{0}, \bM_{t0}^{-1})\le e^{-K_0 T \epsilon_T^2}\mbox{ and }\int (1-\phi_{T,l}) d\N_p(\bm{0}, _{i}^{-1})\le e^{-K_0 T \epsilon_T^2},$$ 
whenever $T^{-1}\sum_{t=1}^T  \sum_{j=1}^p\min\{\Eigen^2_j(\bM_{t0}^{-1/2}(\bM_t -\bM_{t0})\bM_{t0}^{-1/2}), c_0\}>m\epsilon_T^2$. 

If $T^{-1}\sum_{t=1}^T  \sum_{j=1}^p\min\{\Eigen^2_j(\bM_{t0}^{1/2}(\bM_t^{-1}-\bM_{t0}^{-1})\bM_{t0}^{1/2}), c_0\}>m\epsilon_T^2$ holds instead of the last condition, the same conclusion holds.

\end{lemma}

The proofs of these Auxiliary Lemmas are in the Supplementary Section S3.

\section{Simulation}
\label{sec:simulation}
To evaluate the performance of our proposed model, we run four simulation experiments. For all the cases, the sparse marginal precision matrix, $\bOmega_0$, is generated by the following two steps.
\begin{enumerate}
    \item[(1)] Adjacency matrix: Generate three small-world networks, each containing 20 disjoint sets of nodes. Then randomly connect some nodes across the small worlds with probability $q$. The parameters in the small-world distribution and $q$ are set in a way that maintains sparsity within each small-world network.
    \item[(2)] Precision matrix: Generate using G-Wishart with the scale equal to 6 and the adjacency matrix in step (1), and truncate entries smaller than 1 in magnitude.
\end{enumerate}
The simulation settings differ in the time series generation scheme. The specific differences are described at the beginning of each subsection. In the first two simulation schemes, data are generated following the OUT. The third simulation setting is used to investigate the robustness of the OUT framework and considers a VAR(1) model with a pre-specified sparse precision matrix. The final simulation considers univariate series generated from Gaussian processes with Mat\'ern kernels with a small smoothness parameter, yielding less smooth trajectories. We compare our methods with the sparse precision matrix estimates from the Gaussian graphical model (GGM) and Gaussian copula graphical model (GCGM) using the R package {\tt BDgraph} \citep{BDgraphR}.

For \(a=3\), the nugget coefficient \(T^{-a}\) is at most \(40^{-3}=1.5625\times10^{-5}\) over the reported sample sizes. Numerical comparisons show that including or omitting the nugget yields identical results to the reported decimal places. We therefore omit the nugget in the simulations for numerical simplicity.

\subsection{Simulation setting 1}

We follow the OUT model to generate multivariate time series data, and the associated univariate stationary series are generated from Gaussian processes with an exponential kernel. These Gaussian processes only differ in the range parameter, which is uniformly generated from (0, 10).

\begin{table}
\centering
\begin{tabular}{|r|rrr|rrr|rrr|}
\hline
    $T$&\multicolumn{3}{|c|}{5\% Non-zero} & \multicolumn{3}{|c|}{10\% Non-zero} &\multicolumn{3}{|c|}{25\% Non-zero}\\
    \hline
  \hline
 & OUT & GCGM & GGM & OUT & GCGM & GGM & OUT & GCGM & GGM \\ 
  \hline
40 & 1.77 & 1.85 & 15.76 & 2.28 & 3.49 & 16.47 & 7.41 & 8.03 & 9.15 \\ 
  70 & 1.01 & 1.84 & 11.55 & 2.12 & 2.59 & 12.47 & 7.12 & 8.42 & 8.90 \\ 
  100 & 0.89 & 2.02 & 9.85 & 1.73 & 2.08 & 11.23 & 6.14 & 8.39 & 9.84 \\ 
   \hline
\end{tabular}
\caption{Estimation MSE for a precision matrix of dimension $60\times 60$ when $\bZ_{\ell}$s are generated from a Gaussian process with an exponential kernel.}
\end{table}

\subsection{Simulation setting 2}
Here we again follow the OUT model and generate the univariate series from ARMA(1,1) model as $z_{i,t}= \phi_i z_{i,t-1}+\theta_i \varepsilon_{i, t-1}+\varepsilon_{i, t}$ with $\varepsilon_{i, t-1}\sim\Normal(0, \sigma_e^2)$ with $\sigma_{i,e}^2=({1-\phi_i^2})/({1+2\theta_i\phi_i+\theta_i^2})$.
We generate $\theta_i,\phi_i\sim\Unif((0.9,1)\cup(-1, -0.9))$ to obtain a time series with strong dependence.

\begin{table}[htbp]
\centering
\caption{Estimation MSE for a precision matrix of dimension $60\times 60$ when $\bZ_{\ell}$s are generated from a causal ARMA(1,1) process.}
\begin{tabular}{|r|rrr|rrr|rrr|}
\hline
    $T$&\multicolumn{3}{|c|}{5\% Non-zero} & \multicolumn{3}{|c|}{10\% Non-zero} &\multicolumn{3}{|c|}{25\% Non-zero}\\
    \hline
  \hline
 & OUT & GCGM & GGM & OUT & GCGM & GGM & OUT & GCGM & GGM \\ 
  \hline
40 & 1.64 & 2.00 & 13.90 & 1.54 & 1.95 & 11.45 & 8.37 & 9.60 & 14.65 \\ 
  70 & 1.43 & 1.84 & 13.51 & 1.46 & 1.80 & 9.85 & 7.62 & 8.65 & 11.79 \\ 
  100 & 1.18 & 1.60 & 10.87 & 1.45 & 1.52 & 8.61 & 7.16 & 8.14 & 9.58 \\ 
   \hline
\end{tabular}
\end{table}

\subsection{Simulation setting 3}
In this section, we consider a forward version of the backward algorithm given in \cite{roy2017constrained} to generate the data from a causal VAR(1) such that the marginal precision matrix at each time point is fixed at a pre-specified matrix $\bOmega$. The specific algorithm is compiled below to obtain the VAR coefficient ($\bPhi$) and innovation variance ($\bSigma_e$) from a prespecified precision $\bOmega$ and two additional auxiliary parameters $\bK$ and $\bQ$, where $\bK$ is a matrix with unrestricted entries and $\bQ$ is an orthogonal matrix.

\RestyleAlgo{ruled}
\begin{algorithm}[H]
   1. $\bGamma(0)^{-1} = \bC_0^{-1} = d_0^{-1}= \bOmega$
  
  2. $\bC_1^{-1} = \bC_0^{-1}+\bK\bK^T.$ 
  
     3. $\bSigma_e = \bC_0-\bC_0\bK\bT\bK^T\bC_0.$ where $\bT=(\bI+\bK^T\bC_0^{-1}\bK)^{-1}$

      4. $\bK_1 = \bC_0\bK\bT^{1/2}$
    
      5. $\bGamma(1) = \bK_1\bQ d_0^{1/2}.$
    
      6. $\bPhi=\bGamma(1)\bOmega$
    \caption{Get stationary VAR coefficients from a modified \cite{roy2017constrained}'s parameterization.}
    \label{algo1}
\end{algorithm}

The entries in the $\bK$ matrix related to this algorithm are generated from Normal(0,1) and the orthogonal matrix $\bQ$ is generated using {\tt randortho} of R package {\tt pracma} \citep{pracmaR}. Finally, we generate the data using this $\bPhi$ and $\bSigma_e$.

\begin{table}[htbp]
\centering
\caption{Estimation MSE for a precision matrix of dimension $60\times 60$ when the data is generated from VAR(1).}
\begin{tabular}{|r|rrr|rrr|rrr|}
\hline
    $T$ &\multicolumn{3}{|c|}{5\% Non-zero} & \multicolumn{3}{|c|}{10\% Non-zero} &\multicolumn{3}{|c|}{25\% Non-zero}\\
    \hline
  \hline
  & OUT & GCGM & GGM & OUT & GCGM & GGM & OUT & GCGM & GGM \\ 
  \hline
 40 & 1.86 & 2.29 & 8.83 & 3.61 & 4.54 & 17.24 & 12.36 & 15.14 & 16.01 \\ 
  70 & 1.78 & 2.29 & 6.82 & 3.12 & 4.11 & 16.61 & 11.91 & 15.50 & 15.91 \\ 
  100 & 1.57 & 2.16 & 4.33 & 2.95 & 3.23 & 13.09 & 10.25 & 14.79 & 14.97 \\ 
   \hline
\end{tabular}
\end{table}

\subsection{Simulation setting 4}
Here we again follow the OUT framework to generate the data. To increase roughness in the data, we consider Mat\'ern kernels with smoothness parameters drawn from $\Unif(0,1)$ and ranges drawn from $\Unif(0,10)$. 
It also allows the marginal variances of the latent time series to be 1.




\begin{table}[htbp]
\centering
\caption{Estimation MSE for a precision matrix of dimension $60\times 60$ when $\bZ_{\ell}$s are generated from Mat\'ern processes with smoothness bounded in $(0,1)$.}
\begin{tabular}{|r|rrr|rrr|rrr|}
\hline
    $T$&\multicolumn{3}{|c|}{5\% Non-zero} & \multicolumn{3}{|c|}{10\% Non-zero} &\multicolumn{3}{|c|}{25\% Non-zero}\\
    \hline
  \hline
 & OUT & GCGM & GGM & OUT & GCGM & GGM & OUT & GCGM & GGM \\ 
  \hline
40 & 3.63 & 3.61 & 6.78 & 6.07 & 7.39 & 7.27 & 15.46 & 15.51 & 17.66 \\ 
  70 & 2.55 & 3.33 & 5.91 & 5.64 & 6.17 & 6.76 & 10.85 & 12.20 & 16.19 \\ 
  100 & 2.52 & 3.05 & 5.76 & 4.53 & 5.05 & 4.97 & 9.61 & 11.12 & 10.84 \\ 
   \hline
\end{tabular}
\end{table}

Across the four cases above, we find that precision matrix estimation accuracy is higher when the OUT model is used. In case 4, the dependency is the lowest among all cases, and all three methods perform similarly, with OUT being marginally better than the others. While the first two settings adhere to the OUT framework for data generation, the last does not. Thus, it is expected that OUT may outperform other methods in the first two settings but not necessarily in the third setting. However, in all four cases, the proposed method outperforms the two alternatives. This may be because the proposed method accounts for temporal dependence more effectively than the other two methods.

\section{Analysis of GDP industry output data}
\label{sec:data}

We analyze the graphical association among several components of GDP output.
This is a quarterly dataset from 2010 to 2019, downloaded from \url{bea.gov}.
There are thus 40 time points.
Our primary focus is to analyze the associations between durable and non-durable goods. We consider three cases: 1) a model with durable and non-durable goods only (19 variables in total), 2) a model with durable and non-durable goods and non-service type variables (49 variables in total), and 3) a model with all the macroeconomic variables excluding the variables under the government category (65 variables in total). As a pre-processing step, we first take the logarithm and then apply order-1 differencing to remove any time trend before fitting the different models.
The processed data is expected to have a mean of zero.

\begin{figure}[htbp]
\caption{Estimated graphical connections among the durable and non-durable macroeconomic components under different models.}
\centering
\subfigure[Small ($p=19$)]{\includegraphics[width = 0.9\textwidth]{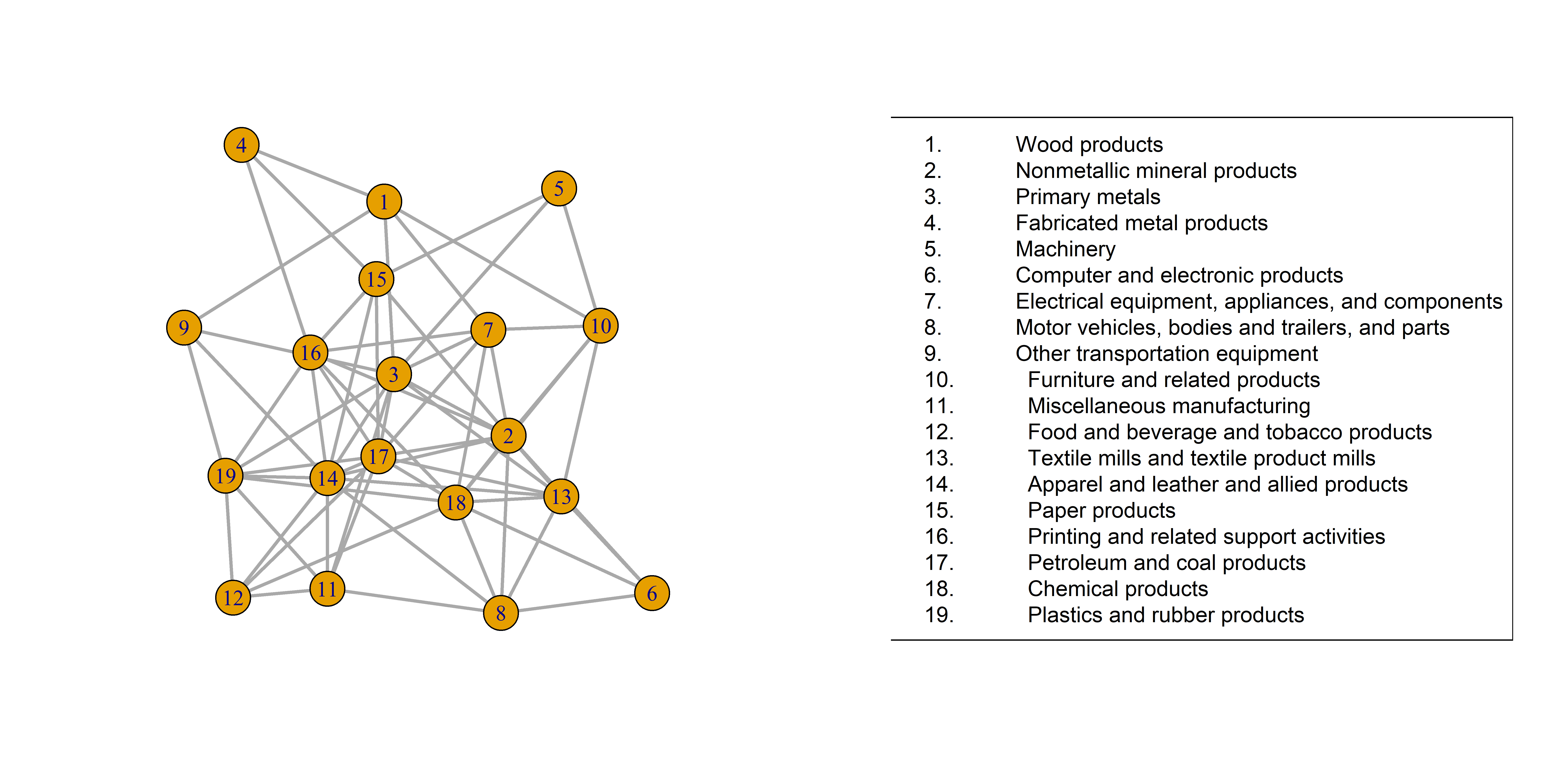}}
\subfigure[Medium ($p=49$)]{\includegraphics[width = 0.45\textwidth]{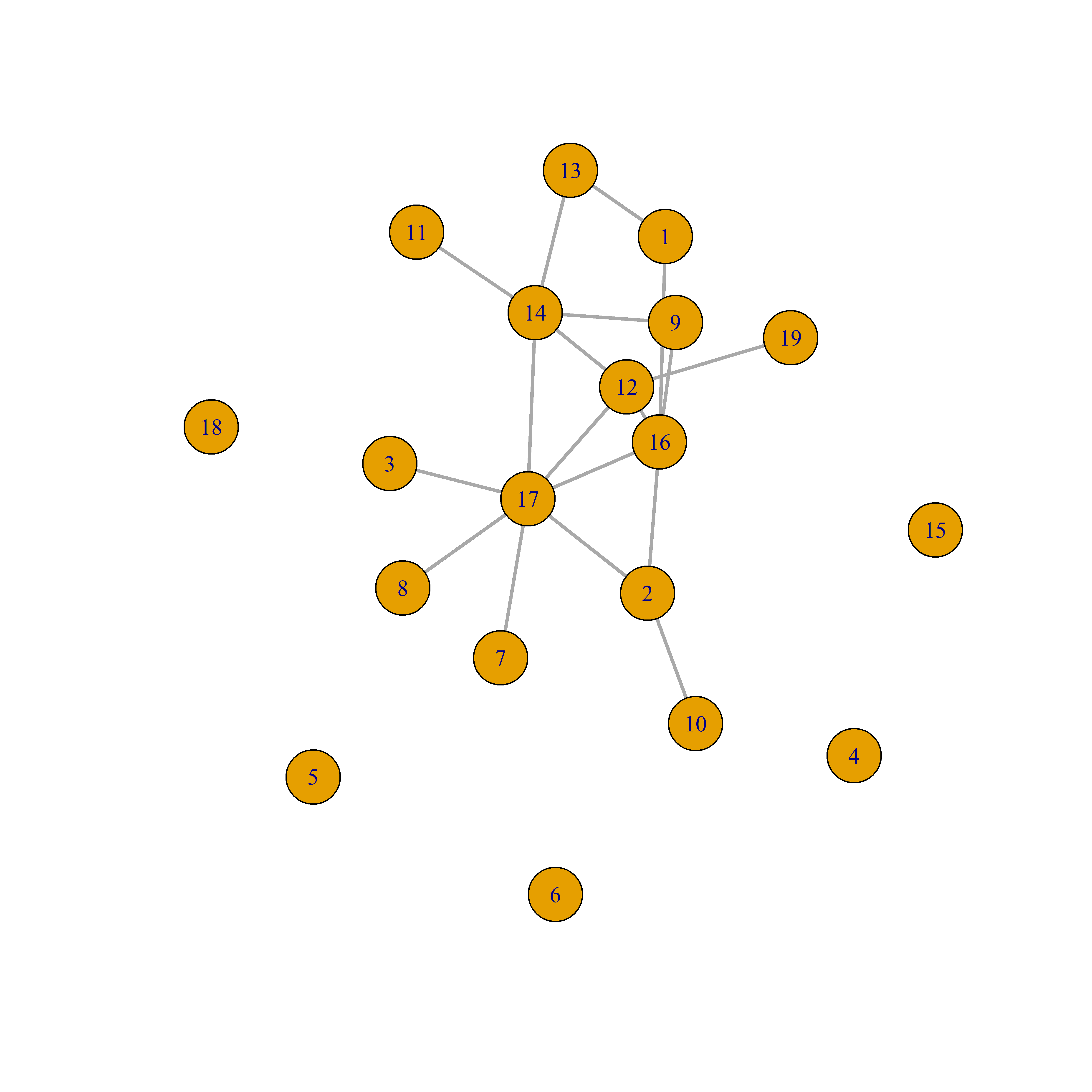}}
\subfigure[Large ($p=66$)]{\includegraphics[width = 0.45\textwidth]{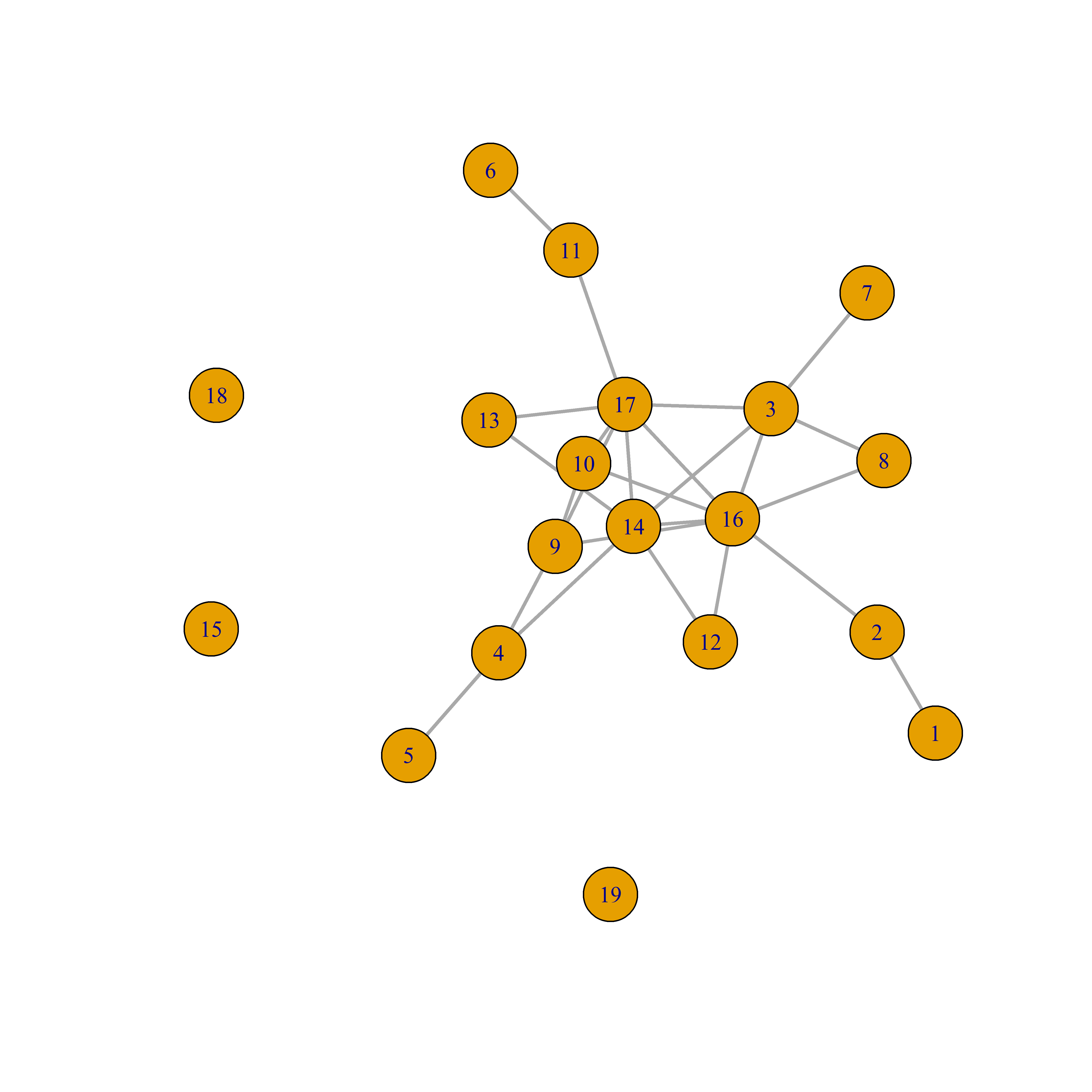}}
\label{fig::real}
\end{figure}

To identify the graphical dependence, we first estimate $\hat{\bOmega}=\hat{\bSigma}^{-1}$.
Subsequently, we scale the matrix and identify entries as significant if they exceed a given absolute threshold. We fix the threshold at 0.1 for the present analysis. 
The edges between the pair of nodes associated with a significant value of ${\hat{\bOmega}}$ are considered as the existing edges in the estimated graph.
Figure~\ref{fig::real} illustrates the estimated sub-graph structures of the 19 durable and non-durable goods for all three cases. As we move from the small to the large model with 65 variables, some connections persist across all three cases. A few pairs are disconnected in a large model. Hence, these are the pairs of variables that become conditionally independent as more variables are added to the analysis. Furthermore, we find that some of the connections are reconstructed in the large case, whereas they are missing in the medium case. This indicates that the newly added macroeconomic variables may have implicit associations with durable and non-durable goods.

Although analyzing graphical dependencies is our primary focus, we also evaluate OUT's predictive performance relative to VAR models to assess its flexibility as a general framework for multivariate time series analysis.
The results are in Tables~\ref{tab:pred1} and \ref{tab:pred2}.

\section{Discussion}
In this paper, we propose a novel multivariate time-series model that is amenable to estimating the conditional independence structure in multivariate data at a given time point. We develop a Bayesian inference method and computationally efficient posterior sampling steps. We further establish posterior contraction rates under the asymptotic regime where the number of time points and the dimension of the multivariate data go to infinity. The numerical performance of the proposed method is illustrated on both synthetic data and the components of GDP. The marginal conditional independence structure provides several interesting interpretations of the GDP data.

Although we take a semiparametric approach to modeling the univariate time series, it would be interesting to consider parametric models, such as AR, MA, or ARMA, for the latent univariate processes. We may then be able to make more explicit inferences about features associated with the temporal dynamics. In addition, one may incorporate a time-varying error variance in the univariate time series when learning about stochastic volatility is of interest. For example, the series can be modeled using ARCH or GARCH frameworks. This may be appropriate while studying multivariate financial time series. Even the parameters in the Cholesky structure may also be assumed to vary over time, leading to alternative modeling formulations. Such extensions would be particularly useful for capturing time-varying brain network dynamics. Another future direction is to adopt the proposed OUT model to address other types of structure-learning problems in multivariate time series, such as learning a directed acyclic graph (DAG) or a dynamic graphical relation. 
In the case of a dynamic graphical model, full stationarity is not possible. However, local stationarity may be incorporated alongside the general OUT formulation.
Further structures can be imposed on the Cholesky parametrization. Future work will also consider establishing graph selection consistency results. 
If we dispose of the stationarity assumption, one or both of $\bL$ and $\bD$ may be varied over time to get time-varying marginal precisions, thereby producing time-varying graphical dependencies. This may be an interesting future direction to pursue.
From a computational perspective, the Metropolis--Hastings (MH) and Langevin Monte Carlo (LMC) sampling steps are computationally intensive, but they enable sampling from the full posterior distribution. This also enables a principled graph-identification procedure by quantifying connection strengths via posterior inclusion probabilities. Nonetheless, it is important to develop more computationally efficient approaches to inference.
The proposed method is developed for continuous time series.
From a practical perspective, extending to mixed-type data will be useful.
One may also consider modeling the univariate time series by a standard parametric model such as an AR or ARMA process.

\section*{Funding}
The research is partially supported by the National Science Foundation collaborative research grants DMS-2210280 (Subhashis Ghosal) / 2210281 (Anindya Roy) / 2210282 (Arkaprava Roy).

\section*{Data availability}
The data is publicly available, and the code is attached to the submission.


\subsection*{Additional real data results}
\label{app:result}

When there are several variables, the sparse VAR is preferable to the traditional VAR.
Specifically, we compare the predictive performance of our proposed model with sparse VAR which is implemented using R package {\tt sparsevar} \citep{sparsevarR}. The sparse VAR employs a penalized least-squares method with an elastic net penalty. We also considered alternative penalties; however, the changes in the estimates were minimal. We fit the sparse-VAR model for 5 choices of the lag parameter. We fit the VAR model for several lag lengths. 
In Tables~\ref{tab:pred1} and \ref{tab:pred2}, we find that the forecast accuracy of our model is always comparable with VAR, despite VAR allowing greater flexibility in terms of the autocorrelation structure. 
In contrast to VAR, the OUT estimates are guaranteed to be stationary by construction.
It is, however, important to note that the superiority in prediction should be tested statistically rather than in terms of MSE numbers alone.
Here, we do not examine this due to the inherent complexities of both sparseVAR and the OUT class of models.

\begin{table}[ht]
\centering
\caption{One-step ahead prediction MSE comparison between OUT and different sparse-VAR models for all economic variables.}
\begin{tabular}{rrrrrrr}
  \hline
 & VAR-1 & VAR-2 & VAR-3 & VAR-4 & VAR-5 & OUT \\ 
  \hline
Small ($p=19$) & 1.35 & 0.45 & 0.41 & 0.42 & 0.51 & {\bf 0.24} \\ 
  Medium ($p=49$) & 0.72 & {\bf 0.57} & 0.62 & 0.77 & 0.76 & 0.77 \\ 
  Large ($p=65$) & 0.55 & 0.43 & 0.46 & 0.59 & 0.58 & {\bf 0.42} \\ 
   \hline
\end{tabular}
\label{tab:pred1}
\end{table}

\begin{table}[ht]
\centering
\caption{One-step ahead prediction MSE comparison between OUT and different sparse-VAR models for the subset of durable variables only.}
\begin{tabular}{rrrrrrr}
  \hline
 & VAR-1 & VAR-2 & VAR-3 & VAR-4 & VAR-5 & OUT \\ 
  \hline
Small ($p=19$) & 1.35 & 0.45 & 0.41 & 0.42 & 0.51 & {\bf 0.24} \\  
   Medium ($p=49$)  & 0.27 & {\bf 0.25} & 0.28 & 0.26 & 0.34 & 0.30\\ 
  Large ($p=65$) & 0.27 & {\bf 0.25} & {\bf 0.25} & 0.26 & 0.34 & 0.28 \\ 
   \hline
\end{tabular}
\label{tab:pred2}
\end{table}

\section{Proof of the main theorems}
\label{sec:theorem proofs}

\begin{proof}[Proof of Theorem 1]
As in the proof of Theorem~8.19 of \cite{ghosal2017fundamentals}, (i) implies the evidence lower bound of the form 
$    \int (q_\theta^{(T)}/q_{\theta_0}^{(T)}) d\Pi(\theta)\ge e^{-D T\epsilon_T^2}$ 
with $p_{0}^{(T)}$-probability tending to one, for some constant $D>0$. 

By H\"older's inequality, 
$$\int \phi_{T,j} p_{\theta_0}^{(T)}  \le \big( \int \phi_{T,j} q_{\theta_0}^{(T)} \big)^{(b-1)/b}\big(\int (p_{0}^{(T)}/q_{\theta_0}^{(T)}\big)^b  q_{\theta_0}^{(T)})^{1/b}\le 
e^{-K' T\epsilon_T^2}G_T^{1/b} .$$
Now, for any $\theta\in \Theta_{T,j}^*$ such that $d_T(\theta,\theta_0)>\epsilon$, 
$$\E \int (1-\phi_{T,j}) \frac{q_{\theta}^T}{q_{\theta_0}}  \le \big( \int (1-\phi_{T,j}) q_\theta^{(T)} \big)^{(b-1)/b}\big(\int (p_{0}^T/q_{\theta_0}^{(T)}\big)^b  q_\theta^{(T)})^{1/b}\le 
e^{-K' T\epsilon_T^2}G_T^{1/b} ,$$
where $K'=K(b-1)/b$.

Now the proof can be completed by standard arguments by aggregating the tests to $\phi_T=\max\{ \phi_{T,j}: j=1,\ldots,N_T\}$, since by the condition $\log G_T\lesssim T\epsilon_T^2$, the testing error rates are not affected. The argument goes through even though the more general Condition (iii) is assumed instead of the standard entropy condition for the sieve.

\end{proof}

\begin{proof}[Proof of Theorem 2]
We apply Theorem 1 with $p_0^{(T)}=p_{\btheta_0}^{(T)}$, the true density of the observations under the true parameter $\btheta_0$ for $\btheta$ and $q_{\btheta}^{(T)}$ standing for the distribution of the observations under Whittle's approximate distribution of a stationary time series. To verify Condition (i) of it, let $\bV_t$ stand for the dispersion matrix of $\bW_{*t}=(\bW_{jt}: j=1,\ldots,p)$ under the true distribution in the time domain for $t=1,\ldots,T$. For $j=1,\ldots,p$, using  Proposition 4.5.2 of \cite{brockwell1991time}, we have $\|\bF^T\bGamma_{j,T}\bF-\bLambda_{j,T}\|_{\infty}\lesssim T^{-1}$, where $\bGamma_{j,T}$ is the true covariance in time-domain of $j$th univariate time-series and $\bF$ is the Fourier basis matrix. Due to the assumed independence over $j$, for the marginal dispersion $\bV_t$ will be a diagonal submatrix of $\Diag(\bF^T\bGamma_{1,T}\bF,\ldots,\bF^T\bGamma_{p,T}\bF)$. We thus have $\|\bV_t-\bS_t\|_\infty\lesssim T^{-1}$ and so  $\Frob{\bV_t-\bS_{t0}}\lesssim \sqrt{p} T^{-1}$. The true dispersion matrix of $\bX_t$ is therefore $\bJ_t:=\bOmega_0^{-1/2} \bU_0 \bV_t \bU_0\trans \bOmega_0^{-1/2}$ and it satisfies $\Frob{\bJ_t- \bM_{t0}^{-1}}\le \op{\bOmega_0^{-1}}\op{\bU_0}^2 \Frob{\bV_t-\bS_{t0}}\lesssim \sqrt{p} T^{-1}$. 
Using (3.4), the actual true expectation of the log-likelihood ratio $\log q_T(\gamma_0,\bX)- \log q_T(\gamma,\bX)$ is given by 
\begin{align*}
\lefteqn{-\frac12\sum_{t=1}^T (\log\det \bM_t-\log \det \bM_{t0})+\frac12 \sum_{t=1}^T \tr ((\bM_t-\bM_{t0}) \bJ_t)}\\
&&=\frac12 \sum_{t=1}^T [-\log\det (\bI_p+ \bR_t)+  \tr (\bR_t)] + \tr [(\bM_t-\bM_{t0}) (\bJ_t-\bM_{t0}^{-1})],
\end{align*}
where $\bR_t=\bM_{t0}^{-1}(\bM_t-\bM_{t0})$. Using the Cauchy-Schwarz inequality for matrices, namely $|\tr(\bA \bB)|\le \Frob{\bA}\Frob{\bB}$, it follows that  
$$\tr [(\bM_t-\bM_{t0}) (\bJ_t-\bM_{t0}^{-1})]\le \Frob{\bM_t-\bM_{t0}} \times \sqrt{p^2}T^{-1}\le \epsilon_T^3\le \epsilon_T^2,$$ 
whenever $\max_{i} \Frob{\bM_t-\bM_{t0}}\le \epsilon_T$. Thus, the last term in the display can be bounded by $T\epsilon_T^2$.  With $\rho_1,\ldots,\rho_p$ standing for the eigenvalues of $\bR_t$, the expression in the sum inside the first term can be written as $-\log (1+\rho_j)+\rho_j\le \rho_j^2$ whenever $|\rho_j|\le 1/2$. The condition is ensured if $\max_t \Frob{\bM_t-\bM_{t0}}\le \epsilon_T$. Hence, the expression in the display can be bounded by a constant multiple of $T\epsilon_T^2$ if $\max_t \Frob{\bM_t-\bM_{t0}}\le \epsilon_T$. The negative logarithm of the prior probability of the last event is bounded by a constant multiple of $T \epsilon_T^2$ by Lemma 6. This completes the verification of Condition (i). 

For Conditions (ii), (iii), and (iv) of Theorem 1, take the sieve $\mathcal{M}_T$ defined in Lemma 8. On this sieve, the nugget gives $\bOmega\succeq T^{-a}\bI_p$ and hence $\op{\bOmega^{-1}}\le T^a$, while item (i) gives $\op{\bOmega}\le (1+\sqrt{s_T}\,b_T)^2b_T^2+T^{-a}$. Thus, the operator norms entering the test and covering arguments grow at most polynomially in $T$; the remaining details are given in Lemma 8. Finally, by Lemma 1, the logarithm of the integral in (4.3) is bounded by a multiple of $p\lesssim T \epsilon_T^2$. This completes the proof of Assertion (1) for the metric $d$ given by 
$d^2(\bgamma,\bgamma_0)= T^{-1} \sum_{t=1}^T\sum_{j=1}^p \min(\rho_{jt}^2,c_0)$, with $\rho_{jt}$ the $j$th eigenvalue of $\bM^{-1/2}_{t0}(\bM_t-\bM_{t0})\bM^{-1/2}_{t0}$. Thus, Assertion (1) holds. 


The proof of Assertion (2) is similar using the second part of Lemma 8.
\end{proof}

\begin{proof}[Proof of Theorem 3]
It follows from the proof of Lemma 2 that, if all eigenvalues of the matrix $\bM_{t0}^{-1/2}\bM_t\bM_{t0}^{-1/2}$ lie within a fixed compact set not depending on $\bM_t$, $t=1,\ldots,T$, then the minimum operation appearing in the estimates (4.6) and (4.8) can be removed, and consequently also in Lemmas 7 and 8 and ultimately in Theorem 2, by choosing $c_0>0$ sufficiently large. This immediately gives Assertions (a) and (b). The condition holds under the stated assumption that the diagonal entries of the Cholesky decomposition and spectral densities lie within two known positive numbers.

To prove (c), we observe that (b) implies that 
$$\Frob{T^{-1}\sum_{t=1}^T \bM_t^{-1} - T^{-1}\sum_{t=1}^T \bM_{t0}^{-1}}\le m_T \epsilon_T $$ with posterior probability arbitrarily close to one under the true distribution. Due to the integrability condition, we have that uniformly for all spectral densities $f_1,\ldots,f_p$ taking values in a compact interval, $T^{-1}\sum_{i=1}^T |f_j (\omega_{k(i)})-\int f_j(\omega)d\omega|=O(T^{-1})$. Together with the identifiability condition that $\int f_j(\omega) d\omega=1$ for $j=1,\ldots,p$, uniformly for all spectral densities $f_1,\ldots,f_j$ taking values in that  compact interval, 
\begin{align}
   \Frob{ T^{-1}\sum_{t=1}^T \bS_t -\bI_p}=O(\sqrt{p}/T), 
   \label{eq:average Si}
\end{align}
resulting in the convergence 
\begin{align*}
  \Frob{ T^{-1}\sum_{t=1}^T \bM_t^{-1}-\bSigma} &=\Frob{\bOmega^{-1/2}\bU   (T^{-1}\sum_{t=1}^T \bS_t-\bI_p) \bU\trans \bOmega^{-1/2}} 
  \\
  &\le \op{\bOmega^{-1}}  \Frob{ T^{-1}\sum_{t=1}^T \bS_t -\bI_p},
\end{align*}
which is $O(\sqrt{p}/T)=o(\epsilon_T)$. 
Since the true value also satisfies the above bound, in particular, it follows from the triangle inequality that $\Frob{\bSigma-\bSigma_0} \le m_T (1+o(1))\epsilon_T $ with posterior probability arbitrarily close to one under the true distribution, proving Assertion (c). 

Finally, Assertion (d) is obtained from Assertion (c) using the relation $\bOmega-\bOmega_0=\bSigma^{-1}(\bSigma-\bSigma_0)\bSigma_0^{-1}$ which gives the inequality $\Frob{\bOmega-\bOmega_0} \le \op{\bSigma^{-1}} \Frob{\bSigma-\bSigma_0} \op{\bSigma_0^{-1}}$, and the fact that the closeness of $\bSigma$ and $\bSigma_0$ in the Frobenius norm implies their closeness in the operator norm leading to the conclusion that the eigenvalues of $\bSigma$ lie between two positive numbers since $\bSigma_0$ has this property by Assumption (A1). 
\end{proof}

\section{Proof of the auxiliary results}
\label{sec:lemma proofs}

\begin{proof}[Proof of Proposition 1]
The first relation is immediate from the OUT representation (2.1), the orthogonality of $\bU$ and $\mathrm{D}(\bZ_t)=\bI_p$, since $\mathrm{D}(\bY_t)=\bSigma^{1/2}\bU\bI_p \bU\trans \bSigma^{1/2}=\bSigma$.  

The relation (ii) follows from the OUT representation (2.1) that 
\begin{align}
\label{eq:OUT autocorrelation}
\bGamma(h,t;\bY)=\bSigma^{-1/2}\bSigma^{1/2} \bU\E (\bZ_t \bZ_{t+h}\trans) \bU\trans \bSigma^{1/2}  \bSigma^{-1/2} =\bU\E (\bZ_t \bZ_{t+h}\trans)\bU\trans ,
\end{align}
which is symmetric because $\E (\bZ_t \bZ_{t+h}\trans)$ is a diagonal matrix due to the independence of the component series.

As shown in the proof of (ii), $\bGamma(h,t; \bY)=\bU \Diag (\gamma_j(h,t): j=1,\ldots,p) \bU\trans$, so the rows of the orthogonal matrix $\bU$ are the eigenvectors, which do not vary with $h$ or $t$. It also follows that $\{ \gamma_j(h,t): j=1,\ldots,p\}$ are the set of eigenvalues. If all latent processes are stationary, the eigenvalues depend only on the lag $h$. 

To prove covariance-stationarity (respectively, strict stationarity under the additional Gaussianity assumption) and zero trends, it suffices to check that $\E (\bY_t \bY_{t+h}\trans)$ is free from $t$ if all latent processes $(Z_{j,t}:t=1,2,\ldots)$, $j=1,\ldots,p$ are stationary. This is immediate from the OUT representation since $\E (\bY_t \bY_{t+h}\trans)= \bSigma^{1/2} \bU \bS_h \bU\trans (\bR^{-1})\trans$, where $\bS_h= \E (\bY_t \bY_{t+h}\trans)$.  

Finally, to prove that $\bY_t$ is causal when all $Z_{j,t}$, $j=1,\ldots,p$, are causal. By the assumption, there exist functions $H_{j,t}$, $j=1,\ldots,p$, $t=1,2,\ldots$, such that $Z_{j,t}=H_{j,t}(e_{j,t},e_{j,t-1},\ldots)$ for independent random variables $\{ e_{j,s}: s=t,t-1,\ldots\}$, for all $j=1,\ldots,p$. Then defining $e_t=(e_{j,t}: j=1,\ldots,p)$ and  $\bH_t(e_t,e_{t-1},\ldots)=\bR^{-1} \bU (H_{j,t}(e_{j,t},e_{j,t-1},\ldots): j=1,\ldots,p)\trans$, we have that $\bY_t=\bH_t( e_t,e_{t-1},\ldots)$ for all $t$, proving the assertion.  
\end{proof}

\begin{proof}[Proof of Lemma 1]
First, we show that the $b$th moment of the likelihood ratio is indeed finite. It is sufficient to show finiteness for each $j$, since the density factorizes by independence with respect to $j$. For any $j$, 
\[ \frac{\phi_T(\bW_{j*};\bm{0}, \bF\trans\bGamma_{j,T}\bF)^b}{\phi_T(\bW_{j*};\bm{0}, \bLambda_{j,T})^b} = \frac{\det(\bF\trans\bGamma^{-1}_{j,T}\bF)^{b/2}}{\det(\bLambda^{-1}_{j,T})^{b/2}}  \exp(-\frac{b}{2}\bW_{j*}\trans[\bF\trans\bGamma^{-1}_{j,T}\bF - \bLambda^{-1}_{j,T}]\bW_{j*}). \]
Thus, for the finiteness of the $b$th moment of the likelihood ratio, we would need positive definiteness of 
$  \bB_j = b[\bF\trans\bGamma^{-1}_{j,T}\bF - \bLambda^{-1}_{j,T}] + \bLambda^{-1}_{j,T} = b\bF\trans\bGamma^{-1}_{j,T}\bF - (b-1)\bLambda^{-1}_{j,T}]$.
Since $\bGamma^{-1}_{j,T}$ is the finite Toeplitz form defined based on $f_j$, by the result 5.2 (b) of \cite{GrenanderSzego1984} we the the eigenvalues of 
$b\bF\trans\bGamma^{-1}_{j,T}\bF$ are bounded in the interval $[bc_{2j}^{-1}, bc_{1j}^{-1}]$. Similarly, the eigenvalues of $(b-1)\bLambda^{-1}_{j,T}$ are bound in the interval $[(b-1)c_{2j}^{-1}, (b-1)c_{1j}^{-1}].$ using the fact that for two positive definite matrices $\bA_1, \bA_2$, the difference $\bA_1 - \bA_2$ is positive definite if the minimum eigenvalue of $\bA_1$ is strictly bigger than the maximum eigenvalue of $\bA_2$, a sufficient condition for $\bB_j$ to be positive definite is that 
\begin{align}  
bc_{2j}^{-1} > (b - 1)c_{1j}^{-1}, \quad \mbox{that is}, \quad   
   1 < b <  \frac{c_{1j}^{-1}}{c_{1j}^{-1} - c_{2j}^{-1}}=\frac{c_{2j}}{c_{2j} - c_{1j}}. \end{align}
Thus, with the choice $b=(1+\min_j{c_{2j}}({c_{2j} - c_{1j}})^{-1})/2$, the requirement is met. Also, by the asymptotic refinement theorem for functions of eigenvalues of Toeplitz forms (\cite{GrenanderSzego1984}, page 76), and the error rate of discrete approximation of Riemann integrals, we have  
\begin{align}  
\frac{\det(\bF\trans\bGamma^{-1}_{j,T}\bF)^{b/2}}{\det(\bLambda^{-1}_{j,T})^{b/2}}  = \psi_j + O(T^{-1}), 
\end{align}
where $\psi_j$ is a finite constant defined in the theorem. 
Then, completing the integral, we have the value of the integral as 
\begin{align} 
G_T= \prod_{j=1}^p(\psi_j + O(T^{-1})\exp\{-\log\det(b\bF\trans\bGamma^{-1}_{j,T}\bF\bLambda_{j,T} - (b-1)\bI)\}.
\end{align}
Writing  $\eta_{j,t}$s for the eigenvalues of $\bF\trans\bGamma^{-1}_{j,T}\bF\bLambda_{j,T}$, 
we have 
\begin{align*}
&\log\det(b\bF\trans\bGamma^{-1}_{j,T}\bF\bLambda_{j,T} - (b-1)\bI)=\log\det(b(\bF\trans\bGamma^{-1}_{j,T}\bF\bLambda_{j,T} -\bI) + \bI)\\&\quad=\sum_{t=1}^T \log\{b(\eta_{j,t} - 1)+1\}, 
\end{align*}
where, by the positive definiteness of $\bB_{j}$, $b(\eta_{j,t} - 1) > -1.$
From Lemma A1 of \cite{choudhuri2004contiguity}, we have $|\log\det(\bF\trans\bGamma_{j,T}\bF)-\log\det(\bLambda_{j,T})|= O(1)$. Moreover, Lemma A2 of \cite{choudhuri2004contiguity} implies $\sum_{t}(\eta_{j,t}-1)^2$ and $\sum_{t}(\eta_{j,t}-1)$ are uniformly bounded for all $T$. Hence, it leads to
$|\log\det(b\bF\trans\bGamma^{-1}_{j,T}\bF\bLambda_{j,T} - (b-1)\bI)|= O(1).$
Therefore, we have $ \log G_T \asymp p$.
\end{proof}

\begin{proof}[Proof of Lemma 2]
    Let $\bR=\bOmega_1^{-1/2}(\bOmega_2-\bOmega_1) \bOmega_1^{-1/2}$ and let $\rho_1,\ldots,\rho_p$ be the eigenvalues of $\bR$. 

    The equalities in (4.6) and (4.7) follow from straightforward integration of normal densities. 
    
    To prove the inequality in (4.6), we write the expression as $$\exp\{ \frac14 \sum_{j=1}^p (\log (1+\rho_j)-2\log (1+\rho_j/2))\}.$$ The function $h(x)=\log(1+x)-2\log(1+x/2)$ has a unique maximum $0$ at $x=0$, satisfies $h(x)/x^2\to-1/4$ as $x\to0$, and tends to $-\infty$ as $x\downarrow-1$ or $x\to\infty$. Therefore, for any $c_1>0$, there exists $c_2>0$ such that $h(x)\le-c_2\min(x^2,c_1)$ for all $x>-1$. The constant $c_1$ may be chosen arbitrarily large at the expense of making $c_2$ smaller. Thus, the expression in (4.6) is bounded by 
    $$\exp\{ -c_2 \sum_{j=1}^p \min(\rho_j^2,c_1)\}.$$  
    Finally, applying the Frobenius-operator norm inequality on the relation  $\bOmega_2-\bOmega_1=\bOmega_1^{1/2}\bR \bOmega_1^{1/2}$, we get the estimate $\Frob{\bR}\ge \Frob{\bOmega_2-\bOmega_1}/\op{\bOmega_1}$. Substituting in the last inequality, the assertion is established.

    Writing the expression as $\exp\{ \frac12 \sum_{j=1}^p (2\log (1+\rho_j)-\log (1+2\rho_j))\}$ and noting that the function $h(x)=2\log(1+x)-\log (1+2x)$ has a minimum $0$ at $x=0$, $h$ is convex and $h''(x)\le 30$ for $|x|\le 1/3$, we can bound the expression in (4.7) by $\exp\{ 15\sum_{j=1}^p \rho_j^2\}=\exp\{ 15\Frob{\bR}^2\}$ whenever $\Frob{\bR}\le 1/3$. 
    Because $\Frob{\bR}\le \op{\bOmega_1^{-1}} \Frob{\bOmega_2-\bOmega_1}$, the assertion follows by the Frobenius--operator norm inequality.

    The proofs of the relations (4.8) and (4.9) respectively follow the arguments in the proofs of (4.6) and (4.7) by working with $\bR=-\bOmega_1^{1/2} (\bOmega_2^{-1}-\bOmega_1^{-1})\bOmega_1^{1/2}$. 
\end{proof}

\begin{proof}[Proof of Lemma 3]
First, to show that the eigenvalues are bounded above, we uniformly bound the operator norms $\op{\bM_{t0}}$. Observe that 
\begin{align*}
   \op{\bM_{t0}}&=\op{\bSigma_0^{1/2} \bU_0 \bS_t^{-1} \bU_0\trans \bSigma_0^{1/2}}\\&\quad\le \op{\bSigma_0} \op{\bS_t^{-1}} \le \op{\bSigma_0} \max\{ 1/f_{j0}(\omega_{k(i)}): 1\le j\le p\}, 
\end{align*}
which is bounded uniformly in $i$ by Assumptions (A1) and (A3), since $\op{\bU_0}=1$, as all eigenvalues of the orthogonal matrix $\bU_0$ lie on the unit circle in the complex plane. 

To show that the eigenvalues are lower-bounded by a positive constant, it suffices to upper-bound the operator norm of the inverse: 
\begin{align*}
\op{\bM_{t0}^{-1}}&=\op{\bSigma_0^{-1/2} \bU_0 \bS_t \bU_0\trans \bSigma_0^{-1/2}}\\&\quad\le \op{\bSigma_0^{-1}} \op{\bS_t} \le \op{\bSigma_0^{-1}} \max\{ f_{j0}(\omega_{k(i)}): 1\le j\le p\},\end{align*}
which is also bounded uniformly in $i$ by Assumptions (A1) and (A3), and the orthogonality of $\bU_0$. 
\end{proof}

\begin{proof}[Proof of Lemma 4]
For $\bU_1=(\bI_p-\bA_1)(\bI_p+\bA_1)^{-1}$, $\bU_2=(\bI_p-\bA_2)(\bI_p+\bA_2)^{-1}$, observe that 
\begin{align*}
&\Frob{(\bI_p-\bA_1)(\bI_p+\bA_1)^{-1}- (\bI_p-\bA_2)(\bI_p+\bA_2)^{-1}} \\ 
& \quad \le \Frob{\bA_1-\bA_2} \op{(\bI_p+\bA_1)^{-1}} +\op{\bI_p-\bA_2} \Frob{(\bI_p+\bA_1)^{-1}- (\bI_p+\bA_2)^{-1}}\\
& \quad \le \Frob{\bA_1-\bA_2}  \op{(\bI_p+\bA_1)^{-1}} [1+ \op{\bI_p-\bA_2} ] \op{(\bI_p+\bA_2)^{-1}}.
\end{align*}    
As $\bA_1$, $\bA_2$ are skew-symmetric matrices, all its non-zero eigenvalues are purely imaginary. Thus any eigenvalue of $(\bI_p+\bA_1)^{-1}$ or $(\bI_p+\bA_2)^{-1}$, being of the form $(1+ia)^{-1}$, is bounded above in absolute value by $1$. Further, $\op{\bI_p-\bA_2}^2 \le \Frob{\bI_p-\bA_2}^2 \le  p+ p(p-1)B^2$. Substituting these estimates, the conclusion follows. 

\end{proof}

\begin{proof}[Proof of Lemma 5]
By the applications of the matrix norm inequalities, we have 
\begin{align*}
\lefteqn{\Frob{(\bI_p-\bL_1)\bD^2_1 (\bI_p-\bL_1\trans)- (\bI_p-\bL_2)\bD^2_2(\bI_p-\bL_2\trans)} }\\
& \le \Frob{\bL_1-\bL_2} \op{\bD_1}^2 \\
&\quad + \Frob{\bD^2_1-\bD^2_2}
 (\op{(\bI_p-\bL_1)}+\op{(\bI_p-\bL_2)}) \\
&\quad + \op{\bD_2}^2 \Frob{\bL_1-\bL_2}.
\end{align*}    
Since $\bD_1,\bD_2$ are diagonal matrices with entries in $(0,B)$, the result is immediate. 

\end{proof}

\begin{proof}[Proof of Lemma 6]
First we show that for $\epsilon>0$ sufficiently small, if $\Frob{\bOmega-\bOmega_0}\le \epsilon$, $\Frob{\bU-\bU_0}<\epsilon$, and $\max_{1\le t\le T}\Frob{\bS_t-\bS_{t0}}<\epsilon$, then for some constant $C_0$ depending on the true parameters $\bOmega_0$, $\bU_0$ and $f_{10},\ldots,f_{j0}$, the relation $\max \{ \Frob{\bM_t-\bM_{t0}}: 1\le t\le T\}\le C_0 \epsilon$ holds. 

To prove this, we recall the facts that $\op{\bOmega_0}$ and $\op{\bS_{t0}^{-1}}$, $t=1,\ldots,T$, are uniformly bounded by Assumptions (A1) and (A3), and that $\op{\bU_0}=1$. This implies that if $\epsilon>0$ is sufficiently small and if $\Frob{\bOmega-\bOmega_0}\le \epsilon$, $\Frob{\bU-\bU_0}<\epsilon$, and $\max\{ \Frob{\bS_t-\bS_{t0}}: t=1,\ldots,T\}<\epsilon$, then $\op{\bOmega}$ and $\op{\bS_t^{-1}}$, $t=1,\ldots,T$, are also uniformly bounded, and $\op{\bU}=1$ by the orthogonality of $\bU$. Next, we argue that if $\Frob{\bOmega-\bOmega_0}\le \epsilon$, then $\Frob{\bOmega^{1/2}-\bOmega_0^{1/2}}\lesssim \epsilon$. To see this, let $e_1,\ldots,e_p$ be the eigenvalues of the positive definite matrix $\bOmega_0^{-1/2}\bOmega\bOmega_0^{-1/2}$. Then by the Frobenius--operator norm inequality $\Frob{\bA \bB}\le \op{\bA}\Frob{\bB}$ or $\Frob{\bA \bB}\le \Frob{\bA}\op{\bB}$ and Assumption (A1), $\Frob{\bOmega^{1/2}-\bOmega_0^{1/2}}^2\lesssim \Frob{\bOmega_0^{-1/4}\bOmega^{1/2}\bOmega_0^{-1/4}-\bI_p}^2=\sum_{j=1}^p (\sqrt{e_j}-1)^2$. Now if $\Frob{\bOmega-\bOmega_0}$ is sufficiently small, then so is $\Frob{\bOmega_0^{-1/2}\bOmega\bOmega_0^{-1/2}-\bI_p}^2=\sum_{j=1}^p ({e_j}-1)^2$, and hence all eigenvalues $e_1,\ldots,e_p$ lie in a sufficiently small neighborhood of $1$. Therefore,   
$\sum_{j=1}^p (\sqrt{e_j}-1)^2$ can be bounded by a constant multiple of $ \sum_{j=1}^p ({e_j}-1)^2$. Thus  $\Frob{\bOmega^{1/2}-\bOmega_0^{1/2}}\lesssim \Frob{\bOmega-\bOmega_0}$. 

By repeatedly applying the triangle inequality and the Frobenius operator norm inequality, and using the above estimates, we obtain the estimate 
\begin{align}
\Frob{\bM_t-\bM_{t0}} & \lesssim \max\{ \Frob{\bOmega^{1/2}-\bOmega_{0}^{1/2}}, \Frob{\bU-\bU_0}, \Frob{\bS_t^{-1}-\bS_{t0}^{-1}} \} \notag \\
& \lesssim \max\{ \Frob{\bOmega-\bOmega_{0}}, \Frob{\bU-\bU_0}, \Frob{\bS_t^{-1}-\bS_{t0}^{-1}} \}
\label{eq:Frobenius estimate}
\end{align}
uniformly for $t=1,\ldots,T$. 
Then, using the prior independence of the parameters, it will suffice to show that for all sufficiently small $\epsilon>0$, 
\begin{enumerate}
    \item [(i)] $-\log \Pi (\Frob{\bOmega-\bOmega_0}\le \epsilon) \lesssim (p+s) \log (p/\epsilon)$;
    \item [(ii)] $-\log \Pi (\Frob{\bU-\bU_0}\le \epsilon) \lesssim (p+s) \log (p/\epsilon)$;
    \item [(iii)] $-\log \Pi (\max\{\Frob{\bS_t^{-1}-\bS_{t0}^{-1}}:1\le t\le T\} \le \epsilon) \lesssim (p+\epsilon^{-1/\alpha}) \log (1/\epsilon)$ for $K\asymp \epsilon^{-1/\alpha}$.
\end{enumerate}

The bound in (i) was established in the proof of Theorem~4.2 of \cite{shi2021bayesian} for the sparse Cholesky decomposition prior for $\bOmega$ based on independent continuous shrinkage distributions on the entries of $\bL$; the proof using a hard-spike-and-slab distribution on the entries is analogous and is slightly simpler. 

To show the estimate in (ii), in view of Lemma 4, it suffices to show that 
$$-\log \Pi (\Frob{\bA-\bA_0}\le \epsilon/p) \lesssim s \log (p/\epsilon).$$ 
There are $p(p-1)/2$ unrestricted entries in $\bA$, out of which $\bA_0$ has $s$ non-zero entries. A standard probability estimate (see, eg, Example~2.2 of \cite{castillo2012needles}) now establishes the bound when the slab probability decays polynomially in $p$. 

To establish the bound in (iii), since the true spectral densities are bounded away from $0$, it suffices to prove that 
$$-\log \Pi (\max\{ \|f_j-f_{j0}\|_\infty: 1\le j\le p\}  \le \epsilon) \lesssim (p+\epsilon^{-1/\alpha}) \log (1/\epsilon).$$ In view of (4.5), it suffices to use $K\asymp \epsilon^{-1/\alpha}$ terms in the B-spline basis expansion to control the bias within a multiple of $\epsilon$. Hence, using the fact that the normalized B-splines are uniformly bounded by a multiple of $K$, it will be enough to estimate the prior concentration of the $\epsilon/K$ neighborhoods of an $R(p+K)$-dimensional vector of uniformly bounded entries in the Euclidean distance. The resulting estimate is $(\epsilon/K)^{R(p+K)}$. Since $R$ is assumed not to grow with $T$, this leads to (iii). 

The result follows from these estimates, since
\[
\epsilon_T=\max\left\{\sqrt{\frac{p+s}{T}},T^{-\alpha/(2\alpha+1)}\right\}\sqrt{\log T}
\]
satisfies $(p+s)\log (p/\epsilon_T)+(p+\epsilon_T^{-1/\alpha})\log(1/\epsilon_T)\lesssim T\epsilon_T^2$.
 \end{proof}

\begin{proof}[Proof of Lemma 7]
Let $\phi_{T}=\Ind \{ \prod_{t=1}^T \phi_p(\bX_t;\bm{0}, \bM_{t1}^{-1})/\phi_p(\bX_t; \bm{0}, \bM_{t0}^{-1})>1\}$ stand for the likelihood ratio test for testing the simple null $\bX_t\sim \mathrm{N}_p(\bm{0},\bM_{t0}^{-1})$ for all $t=1,\ldots,T$, independently, against the point alternative $\bX_t\sim \mathrm{N}_p(\bm{0},\bM_{t1}^{-1})$, for all $t=1,\ldots,T$, independently. Write
\[
r_{jt}=\Eigen_j\{\bM_{t0}^{-1/2}(\bM_{t1}-\bM_{t0})\bM_{t0}^{-1/2}\}.
\]
Then, by Markov's inequality, Lemma 3, and the first assertion of Lemma 2,
\begin{align*}
\int \phi_{T} \prod_{t=1}^T d\mathrm{N}_p(\bm{0}, \bM_{t0}^{-1}) &\le \prod_{t=1}^T \int \sqrt{\phi_p(\bx_t;\bm{0}, \bM_{t1}^{-1})}\sqrt{\phi_p(\bx_t;\bm{0}, \bM_{t0}^{-1})}  d\bx_t \\
&\quad\le \exp[-c_0' \sum_{t=1}^T \sum_{j=1}^p\min\{r_{jt}^2,c_1\}]
\end{align*}
where $c_0,c_0'>0$ depend on $c_1,c_2$ in Lemma 2 and $b_1,b_2$, and that $c_0$ can be chosen as large as we please at the expense of making $c_2'$ smaller. Hence, the first assertion follows with $K_0=c_0'$.  

By symmetry, it follows that $ \int (1-\phi_{T}) \prod_{t=1}^T d\mathrm{N}_p(\bm{0}, \bM_{t1}^{-1}) \le e^{-cT\epsilon^2}$ as well. Now, by the Cauchy--Schwarz inequality and the estimate in the second assertion of Lemma 2,
\begin{eqnarray*}
    \lefteqn{\int (1-\phi_{T}) \prod_{t=1}^T d\mathrm{N}_p(\bm{0}, \bM_{t2}^{-1}) }\\ 
    && \le \big(\int (1-\phi_{T}) \prod_{t=1}^T d\mathrm{N}_p(\bm{0}, \bM_{t1}^{-1}) \big)^{1/2} \prod_{t=1}^T \big( \int \big( \frac{\phi_p (\bx_t;\bm{0},\bM_{t2}^{-1})}{ \phi_p (\bx_t;\bm{0},\bM_{t1}^{-1})}\big)^2 d\mathrm{N}_p (\bm{0},\bM_{t1}^{-1}) \big)^{1/2}\\
    && \le e^{-c T\epsilon^2/2} \exp\big[{\tfrac{15}{2}}\sum_{t=1}^T \op{\bM_{t1}^{-1}}^2 \Frob{\bM_{t2}-\bM_{t1}}^2\big]. 
\end{eqnarray*}
If $15 \sum_{t=1}^T\op{\bM_{t1}^{-1}}^2 \Frob{\bM_{t2}-\bM_{t1}}^2/2< c_0' T \epsilon^2/4$, which holds if $\max\{\Frob{\bM_{t2}-\bM_{t1}}: 1\le t\le T \}< \epsilon /(C\max_{1\le t\le T}\op{\bM_{t1}^{-1}} )$ for some sufficiently large constant $C>0$,  
the expression in the last display is bounded by $e^{-K_0 T \epsilon^2}$ for $K_0=c_0'/4$.

The last part of the lemma follows from similar arguments using the estimates  (4.7) and (4.9) instead of (4.6) and (4.8). 
\end{proof}

\begin{proof}[Proof of Lemma 8]
The event that $(\bM_t: 1\le t\le T)\not\in \mathcal{M}_T$ can happen only if one of these events occurs: 
\begin{itemize}
\item [(i)] $\|\bL\|_0> L T \epsilon_T^2/\log p$; 
\item [(ii)] $\|\bL\|_\infty>L \sqrt{T}\epsilon_T$; 
\item [(iii)] $\max(d_j,d_j^{-1}) >L \sqrt{T}\epsilon_T$ for some $j=1,\ldots,p$; 
\item [(iv)] $\|\bA\|_0> LT \epsilon_T^2/\log p$; 
\item [(v)] $\|\bA\|_\infty > L \sqrt{T}\epsilon_T$;
\item [(vi)] $\max\{|\xi_{jr}|: 1\le j\le p, 1\le r \le R\} > L \sqrt{T}\epsilon_T$; 
\item [(vii)] $\max\{|\eta_{kr}|: 1\le k\le K, 1\le r \le R\} > L \sqrt{T}\epsilon_T$;
\item [(viii)] $K> L T\epsilon_T^2/\log T$. 
\end{itemize}
We verify that the prior probabilities of all these events can be bounded by $e^{-L' T \epsilon_T^2}$, where $L'>0$ is a constant such that it can be made arbitrarily large by choosing $L$ sufficiently large. 

The prior probability of (i) satisfies the bound since the number of nonzero entries out of $p(p-1)/2$ many entries with a probability of a nonzero polynomially decaying in $p$ has a distribution with tail decaying exponentially, as argued in the proof of Theorem~4.2 of \cite{shi2021bayesian}. The estimate there also establishes the required bound for the prior probability of (ii). The claim for the event in (iii) follows since the inverse-Gaussian prior for $d_j$ has an exponential tail at both zero and infinity. 
The event in (iv) is similar to that in (i). The events in (v), (vi), and (vii) also have an exponentially small prior probability by the tail estimate of a Gaussian distribution. Finally, the Poisson tail of $K$ asserts that the event in (viii) complies with the required bound. 

Write $s_T=LT\epsilon_T^2/\log p$ and $b_T=L\sqrt{T}\epsilon_T$, as in the definition of $\mathcal{M}_T$. Since
$\bOmega=(\bI_p-\bL)\bD^2(\bI_p-\bL)^\top+T^{-a}\bI_p$, on the sieve,
\begin{align*}
\op{\bOmega^{-1}}&\le T^a,\\
\op{\bOmega}&\le (1+\op{\bL})^2\op{\bD}^2+T^{-a}
\le (1+\sqrt{s_T}\,b_T)^2b_T^2+T^{-a}.
\end{align*}
Moreover, the B-spline and link-function bounds give
$\max_{t\le T}\op{\bS_t^{-1}}\lesssim T\epsilon_T^2$ and
$\max_{t\le T}\op{\bS_t}\lesssim K$. Therefore, for some $b>0$,
\begin{align}
\label{eq:M bounds}
\max_{t\le T}\op{\bM_t}
&\lesssim \big\{(1+\sqrt{s_T}\,b_T)^2b_T^2+T^{-a}\big\}T\epsilon_T^2
\lesssim T^b,\\
\max_{t\le T}\op{\bM_t^{-1}}
&\lesssim T^aK\lesssim T^b.
\end{align}

Now we partition the sieve $\mathcal{M}_T$ to show that the required conditions hold. Let $N_0$ be the smallest integer greater than or equal to $ L T \epsilon_T^2/\log p$. There are $N_1\le (p^2)^{N_0}$ many configurations of non-zero locations of the matrix $\bL$ in $\mathcal{M}_T$ and similarly $N_2\le (p^2)^{N_0}$ many configurations of non-zero locations of the matrix $\bA$. Thus $\log N_1, \log N_2\lesssim N_0$. For each such $N_1\times N_2$ configuration, let the non-zero values range over $[-L \sqrt{T}\epsilon_T, L \sqrt{T}\epsilon_T]$. Divide this interval into $N_3\asymp T^B$ equal subintervals, where a sufficiently large $B$ is to be chosen later. The range $(1/(L \sqrt{T} \epsilon_T),L \sqrt{T} \epsilon_T)$ is next subdivided into $N_3$ equal subintervals. 
The range $[-L \sqrt{T} \epsilon_T,L \sqrt{T} \epsilon_T]$ of $\xi_{jr}$, $j=1,\ldots,p$, $r=1,\ldots,R$, is then divided in $N_3$ equal subintervals. Also, for every value $K\le N_0$, the range $[-L \sqrt{T} \epsilon_T,L \sqrt{T} \epsilon_T]$ of the coefficients $\eta_{kr}$, $k=1,\ldots,K$, is also divided in $N_3$ subintervals. Thus for each $N_1 N_2 R K$ possible configuration of non-zero entries of $\bL$, $\bA$ and coefficients $\eta_{kr}$, $k=1,\ldots,K$, $r=1,\ldots,R$, we obtain at most $N_3^{N_0+p+N_0+pR+N_0R}$ pieces, that is the total number of pieces is at most $N\le N_1 N_2 R K N_3^{N_0+p+N_0+pR+N_0R}$. Hence 
$$\log N\le (N_0+p+N_0+pR+N_0 R) \log N_3+\log ( N_1 N_2 R K)\lesssim T \epsilon_T^2.$$ 
In each piece, the maximum discrepancy in $\bOmega$ in terms of the Frobenius distance is bounded by a multiple of $(T \epsilon_T^2)^2 \sqrt{p^2} T^{-B}+T \epsilon_T^2 \sqrt{p} T^{-B}\le T^{-B'}$ in view of Lemma 5, where $B'>0$ can be chosen as large as we please by choosing $B$ large enough. Similarly, by applying Lemma 4, the maximum discrepancy in $\bU$ in terms of the Frobenius distance in each piece is bounded by a multiple of $\sqrt{p^2 (\sqrt{T}\epsilon_T)^2} \sqrt{p^2}T^{-B}\le T^{-B'}$. Finally, we bound the same quantity for each $\bS_t^{-1}$. Since these are diagonal matrices, the Frobenius distance is bounded by $\sqrt{p}$ times the maximum discrepancy in the entries of $\bS_t^{-1}$. Each diagonal entry of $\bS_t$ has a lower bound the same as that of the B-spline coefficients, that is, $\min (\theta_{jk}: 1\le j\le p, 1\le k\le K)$. Within the sieve, by construction, the value of $\theta_{jk}=\Psi(\sum_{r=1}^R \xi_{jr} \eta_{kr})$, where $\Psi(x)=(1+x/(1+|x|))/2$ is the link function, will be at least $\Psi(-R L^2 T \epsilon_T^2)\ge (4R L^2 T \epsilon_T^2)^{-1}$. Hence the maximum discrepancy in each diagonal entry of $\bS_t^{-1}$ within a piece can be at most $ (4R L^2 T \epsilon_T^2)^2 T^{-B}$, and hence the maximum discrepancy in $\bS_t^{-1}$ in terms of the Frobenius distance is bounded by a multiple of $T^{-B'}$. Thus, using \eqref{eq:Frobenius estimate}, within each piece, the maximum discrepancy of $\bM_t$ in the Frobenius distance is bounded by a multiple of $T^{-B'}$ for all $t=1,\ldots,T$. 

From the $l$th piece of the sieve, choose and fix an element $(\bM_t^{(l)}:1\le t\le T)$, and let $\phi_{T,l}$ stand for the likelihood ratio test for testing the point null $\bM_t=\bM_{t0}$ for all $t=1,\ldots,T$, against the alternative $\bM_t=\bM_t^{(l)}$ for all $t=1,\ldots,T$. Fix any sequence $m_T\to\infty$, and collect pieces which are $m_T^2\epsilon_T^2$ away from $(\bM_{t0}: 1\le t\le T)$ in terms of the squared metric 
$$d^2((\bM_t:1\le t\le T), (\bM_{t0}: 1\le t\le T)):=T^{-1} \sum_{t=1}^T \min(\Frob{\bM_t-\bM_{t0}}^2, c_0).$$ 
Since each $\op{\bM_t^{-1}}$ with $(\bM_t:1\le t\le T)\in \mathcal{M}_T$ is bounded by a multiple of $T^{b}$ in view of \eqref{eq:M bounds}, the Type I and Type II error probabilities of $\phi_{T,l}$ are bounded by $e^{-K_0 T m_T^2 \epsilon_T^2}$ provided we make $B'>0$ large enough to ensure that $T^{-B'}<\epsilon_T/(C T^b)$. 

The second part of the theorem with $\bM_t^{-1}$ replacing $\bM_t$ holds by similar arguments using the second part of Lemma 7, by observing the size estimate within the sieve applies equally to both $\bM_t$ and $\bM_t^{-1}$.  
\end{proof}
\bibliographystyle{plainnat}
\bibliography{main}
\end{document}